\documentclass[preprint,
showpacs,preprintnumbers,superscriptaddress,amsmath,amssymb]{revtex4}

\usepackage{graphicx}
\usepackage{dcolumn}
\usepackage{bm,morefloats}
\usepackage{color}

\usepackage{amssymb,bm}

\newcommand{\bea}{\begin{eqnarray}}
\newcommand{\eea}{\end{eqnarray}}

\def\X5sp{{\rm X}_5}
\def\Y3sp{{\rm Y}_3}
\def\Z3sp{{\rm Z}_3}

\def\be{\begin{equation}}
\def\ee{\end{equation}}
\def\bea{\begin{eqnarray}}
\def\eea{\end{eqnarray}}

\def\super3{{}^{(3)}}

\begin{document}

\title{Non-minimally coupled hybrid inflation}

\author{Seoktae Koh}
\email{steinkoh"at"sogang.ac.kr}
\affiliation{
Center for Quantum Spacetime, Sogang University,\\
Shinsu-dong 1, Mapo-gu,
121-742, Seoul, Republic of Korea}%

\author{Masato Minamitsuji}%
\email{masato.minamitsuji"at"kwansei.ac.jp}
\affiliation{
Department of Physics,
Graduate School of Science and Technology, \\
Kwansei Gakuin University, Sanda 669-1337, Japan.
}

\date{\today}

\begin{abstract}
We discuss the hybrid inflation model
where
the inflaton field is nonminimally coupled to gravity.
In the Jordan frame, the potential contains
$\phi^4$ term as well as
terms in the original hybrid inflation model.
In our model, inflation can be classified into
{\it the type (I)} and {\it the type (II)}.
In the type (I), inflation is terminated
by the tachyonic instability of the waterfall field,
while
in the type (II) by the violation of slow-roll conditions.
In our model,
the reheating takes place only at
the true minimum
and even in the case (II)
finally the tachyonic instability occurs
after the termination of inflation.
For a negative nonminimal coupling,
inflation takes place
in the vacuum-dominated region, in the large field region,
or near the local minimum/maximum.
Inflation in the vacuum dominated region
becomes either the type (I) or (II),
resulting in blue or red spectrum of the curvature perturbations,
respectively.
Inflation around the local maximum can be either the type (I) or
the type (II), which results in the red spectrum of the
curvature perturbations,
while it around the local minimum must be the type (I),
which results in the blue spectrum.
In the large field region, to terminate inflation,
potential in the Einstein frame must be positively tilted,
always resulting in the red spectrum.
We then numerically solve the equations of motion
to investigate the whole dynamics of inflaton
and confirm that the spectrum of curvature perturbations
changes from red to blue ones
as scales become smaller.
\end{abstract}

\pacs{11.25.-w, 11.27.+d, 98.80.Cq}

\keywords{Inflation}

\maketitle

\section{Introduction}

Inflation has become one of the constituents in modern cosmology.
In the high energy theory or particle physics,
there are still
many inflationary models
consistent with the current observations
(see e.g., \cite{wmap7} for the seven-year WMAP observations).
Future observations, such as the Planck satellite \cite{planck},
will measure the B mode polarizations in the Cosmic Microwave
Background (CMB)
anisotropy,
which may contain the information of the
primordial gravitational waves,
and will constrain the inflationary models more severely.

The simplest model is
inflation driven by a single field.
However,
in terms of the particle physics models
or string compactifications,
it is more plausible that
there are two or more fields which concern the
inflationary dynamics.
Among these models,
the hybrid inflation is widely studied \cite{hinf,gbw,gb,Riotto:2002yw}.
In the typical model of the hybrid inflation,
the inflaton field $\phi$ rolls down along a valley of $\chi=0$,
where the waterfall field $\chi$ is stable during inflation.
After $\phi$ passes through a critical point $\phi=\phi_c$,
$\chi$ becomes tachyonic and
eventually rolls down toward the true minimum.
Hybrid inflation can be realized,
in the framework of supergravity and superstring theory
(see \cite{rev1} and references therein).

The main purpose of our work is
to study
how the dynamics of hybrid inflation and their observational predictions
are modified due to the nonminimal coupling of inflaton to gravity.
The effects of the nonminimal couplings have been
studied in many publications
(see e.g,
\cite{fm,ms,uzan,sty,Hwang:1996np,tg,koh,kks,bs,py,bks,sf,kaiser,ns,
pallis}).
We classify the possible hybrid inflationary dynamics
and the observational consequences.
In the original hybrid inflation,
inflation takes place in the vacuum dominated region,
but
it is known that
the spectrum of curvature perturbations becomes blue one.
On the other hand,
in a larger field region,
the potential typically becomes super-Planckian.
In this paper, we consider a nonminimal coupling of
the inflaton field to gravity.
A negative nonminimal coupling
suppresses the potential in the Einstein frame
and its value in the large field region
can remain sub-Planckian.
In addition,
to obtain the flat potential in the large field region,
we will add the $\phi^4$ term to the original potential
in the Jordan frame.

This paper is organized as follows.
In Sec. II, we give our
model with
an inflaton field nonminimally coupled to gravity.
In Sec. III, we discuss the inflationary dynamics
and predictions
in the special cases.
Then,
in Sec. IV, we extend our analysis for
the general cases.
The last Sec. V is devoted to give
the brief conclusion and summary.

\section{The model}

\subsection{Model}

We consider two interacting scalar fields.
One of them, denoted by $\phi$, plays the role of inflaton
and is now nonminimally coupled to gravity.
The other field, called the waterfall field and denoted by $\chi$,
has the vanishing amplitude during inflation
and terminates the inflation
because of its tachyonic instability,
after it gets a negative mass square.
The action of our model in the Jordan frame is given by
\bea
S=\int d^4x\sqrt{-g}
\Big[
\frac{1}{2\kappa^2}\Big(1-\xi\kappa^2 \phi^2\Big)R
-\frac{1}{2}\big(\partial \phi\big)^2
-\frac{1}{2}\big(\partial \chi\big)^2
-V(\phi,\chi)
\Big],\label{action_1}
\eea
where $\xi$ and $\kappa^2 = \frac{8\pi}{m_{pl}^2}$ represent
the nonminimal coupling parameter and the gravitational constant,
respectively.
Here, $m_{pl}=1.2\times 10^{19}{\rm GeV}$ is the Planck mass.

With the conformal transformations
\bea
\hat{g}_{\mu\nu} = \Omega^2(\phi) g_{\mu\nu},
\label{eq:conftrans}
\eea
where $\Omega^2(\phi) = 1- \kappa^2 \xi \phi^2$,
it is possible to move to the Einstein frame.
Having the conformal transformation (\ref{eq:conftrans}),
the action (\ref{action_1}) is transformed
as \cite{kaiser}
\bea
S= \int d^4x \sqrt{-\hat{g}} \biggl[ \frac{1}{2 \kappa^2}\hat{R}
- \frac{1}{2}\hat{g}^{\mu\nu}
\hat{\nabla}_{\mu} {\hat \phi} \hat{\nabla}_{\nu}{\hat\phi}
- \frac{1}{2\Omega^{2}} \hat{g}^{\mu\nu}
\hat{\nabla}_{\mu} \chi \hat{\nabla}_{\nu} \chi
- \hat{V}(\hat \phi,\chi) \biggr],
\label{eq:ctaction}
\eea
where
\bea
{\hat \phi} &=& \int F(\phi) d\phi,
\quad
 F(\phi)
=\frac{\sqrt{1-\kappa^2 \xi (1-6\xi)\phi^2}}{\Omega^2},
\quad \hat{V}({\hat \phi}, \chi) = \frac{1}{\Omega^4(\phi)}
V(\phi,\chi).
\eea
For the negative coupling $\xi<0$, we obtain
\bea
\hat \phi
=\frac{1}{\kappa}
\left\{
\sqrt{\frac{1+6|\xi|}{|\xi|}}
{\rm arcsinh}\big(\sqrt{|\xi|(1+6|\xi|)}\kappa\phi\big)
+\sqrt{\frac{3}{2}}
\ln\Big[\frac{-\sqrt{6}|\xi|\kappa\phi+\sqrt{1+|\xi|(1+6|\xi|)\kappa^2\phi^2}}
             {\sqrt{6}|\xi|\kappa\phi+\sqrt{1+|\xi|(1+6|\xi|)\kappa^2\phi^2}}
     \Big]
\right\},
\eea
which gives $\hat\phi\simeq \phi$ for $\phi\to 0$,
and
\bea
\kappa {\hat \phi}
\simeq  \sqrt{\frac{3}{2}}
\ln\Big[1+12|\xi|-2\sqrt{6|\xi|\big(1+6|\xi|\big)}\Big]
+\sqrt{\frac{1+6|\xi|}{|\xi|}}
\ln \Big[
2\sqrt{|\xi|(1+6|\xi|)}\kappa\phi
\Big],
\eea
for $\phi\to\infty$.

As the potential in the Jordan frame,
we consider the following form
\bea
V\big(\phi,\chi\big)
=\frac{\lambda}{4}\big(\chi^2-v^2\big)^2
+\frac{1}{2}m^2\phi^2
+\frac{1}{4}\mu\phi^4
+\frac{1}{2}g^2\phi^2\chi^2,
\label{pot_phi}
\eea
which has the same form as in the ordinary hybrid inflation model
except for the term of $\frac{1}{4}\mu\phi^4$.
Here, $\mu$ is a dimensionless self-coupling constant
assumed to be positive.
In the Einstein frame,
the scalar potential becomes
\bea
{\hat V}(\phi,\chi)=\frac{1}
              {(1-\kappa^2\xi\phi^2)^2}
\Big[
\frac{\lambda}{4}\big(\chi^2-v^2\big)^2
+\frac{1}{2}m^2\phi^2
+\frac{1}{4}\mu\phi^4
+\frac{1}{2}g^2\phi^2\chi^2
\Big].
\label{pot_eins}
\eea
The reasons to add the $\phi^4$ term to the potential
are as follows:
Firstly, in the Jordan frame,
$\phi^4$-theory gives the most general renormalizable theory.
Secondly, in the Einstein frame,
since the denominator of Eq. (\ref{pot_eins})
is the quartic function of $\phi$,
without $\phi^4$ term in the estimator,
as $\phi$ increases the potential approaches zero.
To realize the inflaton in the large $\phi$ region
rolling down toward the origin,
the potential should increase monotonically
and
at least we need $\phi^4$ term in the estimator.
Of course, this term is not sensitive to the inflationary dynamics
in the vacuum-dominated region $\phi\gtrsim 0$.

Before closing this subsection,
we should mention that our model is similar
to that discussed in Ref. \cite{lerner},
except that both the inflaton and the waterfall field
are nonminimally coupled to gravity.
In the model of \cite{lerner}, the waterfall field is assumed to be
the Higgs field, which is somewhat 
close to the Higgs inflation model (see e.g.,
\cite{bs,py,bks,bez,hertz,barbon,blt,lerner2}).
Although in this paper we will not consider Higgs fields,
it also should be noted that
in the Higgs inflation model
radiative corrections may play the important roles
(see e.g., \cite{bks,lerner,hertz}), and 
there are issues about the naturalness and unitarity violation
(see e.g., \cite{bez,hertz,barbon,blt,lerner2}).

\subsection{Inflationary dynamics and observational predictions}

In the rest of the paper, we discuss the inflationary dynamics
in the Einstein frame.
During inflation, since $\phi$ field rolls along
$\chi =0$, it can be treated as a single field inflation
Note that
for a single field case
the observational quantities are conformal
invariant and thus the same as those
in the original Jordan frame (see e.g., \cite{ms,Hwang:1996np,tg,koh}).
In the Einstein frame,
the slow-roll parameters are defined by
\bea
&&\epsilon:=
\frac{1}{2\kappa^2} \left(
\frac{\hat{V}_{,\hat{\phi}}}{\hat{V}}
 \right)^2
=\frac{1}{2\kappa^2}
\Big(\frac{d\phi}{d{\hat \phi}}\Big)^2
\Big(\frac{{\hat V}_{,\phi}}{\hat V}\Big)^2,
\nonumber\\
&&\eta:= \frac{1}{\kappa^2} \frac{\hat{V}_{,\hat{\phi}\hat{\phi}}}{\hat{V}}
=
\frac{1}{\kappa^2}\Big(\frac{d\phi}{d{\hat \phi}}\Big)^2
\frac{1}{\hat V}
\Big[
\frac{d^2{\hat V}}{d\phi^2}
-\frac{\frac{d\hat V}{d\phi}\frac{d^2\hat \phi}{d\phi^2}}
      {\frac{d{\hat \phi}}{d\phi}}
\Big].
\label{sl}
\eea
The slow-roll inflation takes place
as long as the above parameters are less than unity.

Here, we briefly explain the picture of the field dynamics
in the Einstein frame.
In this paper,
we focus on the trajectory $\chi=0$
and do not discuss the motion in $\chi$ direction.
The asymptotic values in
the vacuum-dominated and large field regions
are
$\frac{1}{4}\lambda v^4$ and
$\frac{\mu}{4\kappa^4\xi^2}$, respectively.
The effective theory description is valid,
if inflation takes place at sub-Planckian scale.
Thus, a theoretical bound for the nonminimal
coupling
is obtained as
\bea
\lambda<\Big(\frac{2}{\kappa^2 v^2}\Big)^2,\quad
|\xi|>\mu^{\frac{1}{2}}.\label{qg}
\eea

In terms of the Einstein frame,
inflation can be classified into the following two types,
i.e., {\it the type (I)} or {\it the type (II)}.
In the type (I) inflation,
inflation is terminated by the tachyonic instability
of $\chi$ field.
This instability
appears at $\phi<\phi_c=\sqrt{\frac{\lambda v^2}{g^2}}$.
For $\phi<\phi_c$,
the trajectory deviates from $\chi=0$
and eventually settles down at
either of the global minima $(\phi,\chi)=(0,\pm v)$.
In the type (II) inflation,
it is terminated by the violation of
the slow-roll conditions, at some $\phi=\phi_f$,
where either of $\epsilon$ or $|\eta|$ becomes unity.
The type (I) inflation always requires
that $\phi$ moves from larger to smaller values.
In our model, the only way to realize the reheating
is oscillations of fields
at the true minimum after $\chi$ gets the VEV,
and basically we require the inflaton motion
from larger to smaller field values.

We also introduce the observational quantities
in the linear perturbation theory.
The effects of the $\chi$ fluctuations onto the
observational predictions
are also expected to be negligible
as long as at least
$\chi$ is minimally coupled to gravity
\cite{wf1,wf2,wf3,wf4}.
Thus, in this paper
we apply the single-field, slow-roll approximations.
In the context of the slow-roll approximations,
the amplitude of the curvature and tensor perturbations
are given by
\bea
&&P_{s}=\frac{\kappa^6}{12\pi^2}
   \frac{{\hat V}^3}{{\hat V}_{,\phi}^2}
\Big(\frac{d{\hat \phi}}{d\phi}\Big)^2,
\quad
P_{t}=2\kappa^2\Big(\frac{\hat H}{2\pi}\Big)^2
=\frac{2\kappa^4\hat V}{3\pi^2},
\eea
where $\phi$ is evaluated at
the horizon crossing time $t_{\ast}$, $k=aH|_{t=t_{\ast}}$.
The tensor-to-scalar ratio is given by
\bea
r:=\frac{P_{t}}{P_{s}} = 16\epsilon.
\eea
Finally,
the spectral indices of the curvature and tensor
perturbations are given by
\bea
&&n_{s}-1:=\frac{d\ln P_{s}}{d\ln k}
=-6\epsilon+2\eta,
\quad
n_t:=\frac{d\ln P_t}{d\ln k}=-2\epsilon.
\eea

\section{The cases of $\mu = 0$ or $m=0$}

For simplicity, in this section
we discuss the inflationary dynamics
and predictions in the special cases that $\mu=0$ or $m=0$
in Eq. (\ref{pot_phi}).
In the next section, we will extend our analysis
to the general potential.

\subsection{The case of $\mu=0$}
\label{sect:mu0}

\begin{figure}
\centering
\includegraphics[width=0.8\textwidth]{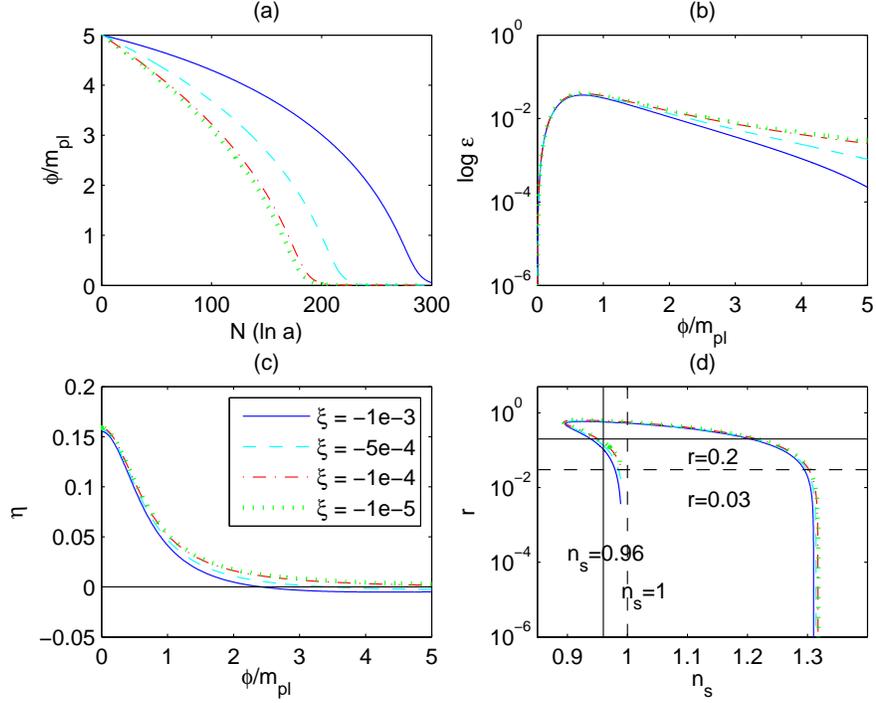}
\caption{ We plot (a) $\phi$ field evolution with the number of efolds $N$
(b) evolution of slow roll parameter $\epsilon$ with $\phi$,
(c) evolution of slow-roll parameter $\eta$ with $\phi$, and (d)
the spectral index of scalar perturbations $n_s$ 
versus tensor to scalar ratio $r$
with $\xi = -10^{-3}, -5\times 10^{-4}, -10^{-4}$ and $-10^{-5}$.
We set $\lambda = g^2 =1, v^2 = 10^{-4}m_{pl}^2,\,\,
m^2 = 10^{-8} m_{pl}^2$.}
\label{figmu0-1}
\end{figure}

In the case of $\mu=0$,
if $0<|\xi|<\frac{m^2}{\kappa^2\lambda v^4}$,
the potential in the Einstein frame has a local maximum at
$\phi = \phi_e$ which is given by
\bea
\phi=\phi_{e}:=\sqrt{\frac{m^2-\kappa^2 |\xi|\lambda v^4}
                 {\kappa^2 |\xi|m^2}}.
\eea
and the potential at $\phi= \phi_e$ takes the value
\bea
{\hat V}(\phi_e,0)=
{\hat V}_e:=\frac{m^4}
         {4\kappa^2|\xi|\big(m^2-2|\xi|\kappa^2\lambda v^4\big)}.
\eea

In the small field region $\phi\gtrsim 0$,
for $|\xi|<\frac{m^2}{\kappa^2 \lambda v^4}$,
$\frac{d\hat V}{d\hat \phi}>0$,
and
for $|\xi|>\frac{m^2}{\kappa^2 \lambda v^4}$,
$\frac{d\hat V}{d\hat \phi}<0$.
In the latter case, since the potential
does not contain the local minimum,
there is no way to terminate the inflation.
Thus, we will not consider the latter case in the
present work.

For $|\xi|<\frac{m^2}{\kappa^2 \lambda v^4}$,
the slow-roll parameters
are given by
\bea
\epsilon
\simeq \frac{8(m^2-|\xi|\kappa^2 \lambda v^4)^2\phi^2}
              {\kappa^2 \lambda^2 v^8},
\quad
\eta\simeq
\frac{8(m^2-|\xi|\kappa^2 \lambda v^4)}
              {\kappa^2 \lambda v^4}>0.
\label{eq:slowroll_m0vac}
\eea
Since as $\phi$ decreases $\epsilon$ decreases,
inflation cannot be the type (II), but be the type (I),
which is the
ordinary hybrid inflation.
The e-folding number is given by
\bea
N =
\int_{{\hat \phi}_f}^{{\hat \phi}_i} 
\frac{\kappa^2 {\hat V}}{{\hat V}_{,\hat\phi}} d\hat \phi
=
\int_{\phi_f}^{ \phi_i} 
\frac{\kappa^2 {\hat V}}{{\hat V}_{,\phi}}
\Big(\frac{d\hat \phi}{d\phi}\Big)^2
 d\phi
= \int_{\phi_f}^{\phi_i} \frac{\kappa d\phi}{\sqrt{2} \epsilon^{1/2}} 
\simeq
\frac{\kappa^2\lambda v^4}{4(m^2-|\xi|\kappa^2\lambda v^4)},
\label{eq:efold_m0vac}
\eea
where $\phi_i$ and $\phi_f$ are the values at the beginning and the
end of inflation, respectively, and we omit the logarithmic factor
at the final expression.
The slow-roll inflation can be realized for
\bea
0<\phi<\phi_i:=\frac{\kappa\lambda v^4}
       {2\sqrt{2}(m^2-\kappa^2|\xi|\lambda v^4)},
\eea
where $\phi_i$ is determined from the condition
that $\epsilon_i =1$.
Since $N\simeq \frac{\kappa\phi_i}{\sqrt{2}}$,
to obtain sufficiently long inflation, 
$\phi_i\gg\frac{1}{\kappa}$.
Also, since $\phi_e>\phi_i$, we obtain $\phi_e\gg \frac{1}{\kappa}$,
which leads to
$$
|\xi|<\frac{m^2}{m^2+\kappa^2\lambda v^4}.
$$
Then, we obtain
\bea
N<\frac{\kappa^2 \lambda v^4}{4m^2}
\Big(1+\frac{\kappa^2 \lambda v^4}{m^2}\Big).
\eea
The requirement $N>50$ gives
$\frac{\kappa^2 \lambda v^4}{m^2}\gtrsim 15$.

In this case,
the amplitude of the curvature perturbations, the tensor-to-scalar ratio,
and the spectral indices of the curvature and tensor perturbations
are given by
\bea
P_{s}&\simeq &\frac{v^{12}\kappa^6 \lambda^3}
    {768 \pi^2\big(m^2-|\xi|\kappa^2\lambda v^4\big)^2\phi_\ast^2},
\quad
r\simeq \frac{128(m^2-|\xi|\kappa^2 \lambda v^4)^2\phi_\ast^2}
       {v^8\kappa^2\lambda^2},
\nonumber\\
n_{s}-1&\simeq &8\big(\frac{m^2}{\kappa^2 \lambda v^4}-|\xi|\big)
 \simeq \frac{2}{N} >0,\quad
n_t\simeq -\frac{16(m^2-|\xi|\kappa^2 \lambda v^4)^2\phi_{\ast}^2}
          {\kappa^2\lambda^2 v^8}\,.
\eea
Thus, the spectrum is blue tilted.


In the region of $\phi>\phi_e$, there is no way to terminate
inflation.
Around the local maximum $\phi\lesssim\phi_e$,
inflation can be either the type (I) or the type (II).
For the tachyonic instability to appear,
we need the condition $\phi_{e}>\phi>\phi_c$,
which leads to
\bea
0<|\xi|<\frac{g^2m^2}{\kappa^2\lambda v^2(g^2v^2+m^2)}.
\eea
The slow-roll parameters are given by
\bea
\epsilon&\simeq&
 \frac{32|\xi|^2\kappa^2
m^2(m^2-|\xi|\kappa^2\lambda v^4)^2
                     (\phi-\phi_{e})^2}
                    {(2m^2-\kappa^2\lambda v^4)^2
\big[
m^2+(m^2-|\xi|\kappa^2\lambda v^4)(1+6|\xi|)
\big]},
\nonumber\\
\eta&\simeq&
-\frac{8|\xi|(m^2-\kappa^2\lambda v^4|\xi|)}
      {m^2+(m^2-|\xi|\kappa^2\lambda v^4)(1+6|\xi|)}<0.
\eea
The slow-roll condition is violated
when $\epsilon=1$.
The e-folding number is given by
\bea
N\simeq \frac{m^2+(m^2-|\xi|\kappa^2\lambda v^4)(1+6|\xi|)}
{8|\xi|(m^2-\kappa^2\lambda v^4|\xi|)}.
\eea
The amplitude of the curvature perturbations and the tensor-to-scalar ratio are given by
\bea
P_{s}
&\simeq &
\frac{(2m^2-\kappa^2|\xi|\lambda v^4)
\big(m^2+(m^2-|\xi|\kappa^2\lambda v^4)(1+6|\xi|)\big)}
{3072|\xi|^3\pi^2(m^2-|\xi|\kappa^2\lambda v^4)(\phi_{\ast}-\phi_{e})^2},
\nonumber\\
r&\simeq &
\frac{512\kappa^2m^2 |\xi|^2
(m^2-|\xi|\kappa^2\lambda v^4)^2(\phi_{\ast}-\phi_{e})^2}
{(2m^2-\kappa^2|\xi|\lambda v^4)^2
\big(m^2+(m^2-|\xi|\kappa^2\lambda v^4)(1+6|\xi|)\big)},
\eea
respectively.
The spectral index of the curvature perturbations is given by
$n_{s}-1\simeq 2\eta\simeq -\frac{2}{N}<0$,
namely red spectrum.

In order to follow the dynamics of the $\phi$ field from the
local maximum to the vacuum dominated region, we perform the
numerical calculations by solving the equations of motion.
For the parameters $\lambda = g^2 = 1,\,\, v^2=10^{-4}\,m_{pl}^2,
\,\, m^2 = 10^{-8}\, m_{pl}^2$, we plot the evolution of the $\phi$ field,
the evolution of the slow-roll parameters $\epsilon$ and $\eta$
and the relation between $n_s$ and $r$ in Fig. \ref{figmu0-1}.
We take into account $\phi_i$ 
being smaller than the local maximum $\phi_e$.
In Fig. \ref{figmu0-1}-(a), we find that enough number of e-folds
can be obtained. The evolution of the slow-roll parameters are
plotted in Figs. \ref{figmu0-1}-(b) and \ref{figmu0-1}-(c).
Theses figures show that the slow-roll conditions are not
violated, 
so inflation should be
terminated through the tachyonic 
instability.
We plot $n_s-r$ relation in Fig. \ref{figmu0-1}-(d). The spectrum
changes from red around the local maximum
to blue in the vacuum dominated region.

\subsection{The case of $m=0$}

\begin{figure}
\centering
\includegraphics[width=0.8\textwidth]{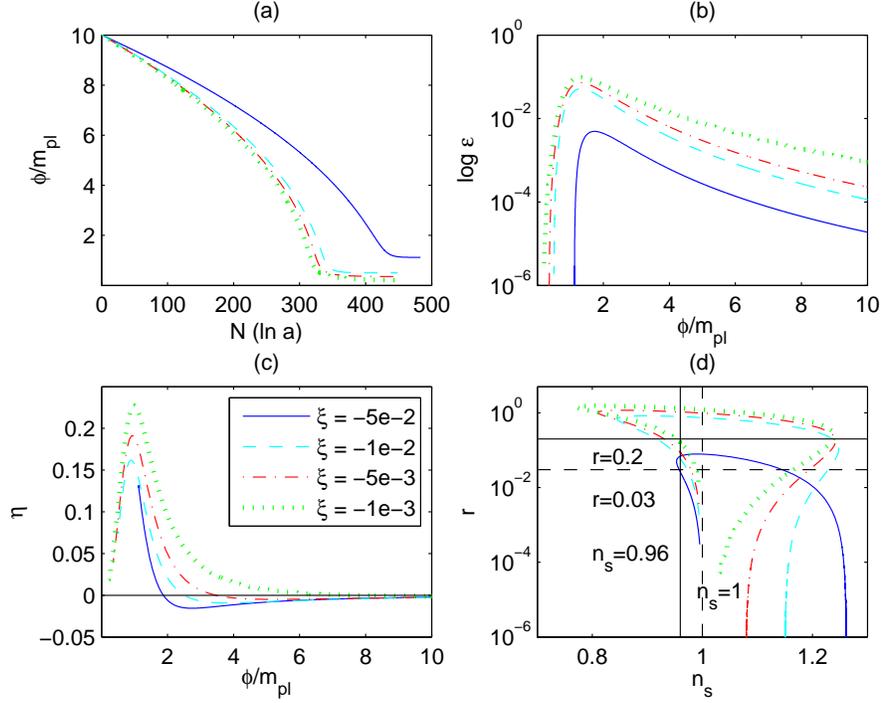}
\caption{ We plot (a) $\phi$ field evolution with the number of efolds $N$,
(b) evolution of slow roll parameter $\epsilon$ with $\phi$,
(c) evolution of slow-roll parameter $\eta$ with $\phi$ and (d)
spectral index of scalar perturbations $n_s$ versus tensor to scalar ratio $r$
with $\xi = -5\times 10^{-2}, -10^{-2}, -5\times 10^{-3}$ and $-10^{-3}$.
We set $\lambda = g^2 =1,\,\, v^2 = 10^{-4} \, m_{pl}^2,\,\,
\mu = 10^{-8}$.}
\label{figm0-1}
\end{figure}


In the case of $m=0$ but $\mu\neq 0$,
along $\chi=0$
the potential
has the local minima at $\phi = \phi_e$,
\bea
\phi=\phi_{e}:=
\sqrt{\frac{\kappa^2\lambda v^4 |\xi|}{\mu}},
\label{past}
\eea
and the potential at $\phi=\phi_e$
takes the nonzero minimum value
\bea
{\hat V}(\phi_e,0)=
{\hat V}_e:=\frac{v^4\lambda\mu}
{4(\kappa^4\lambda v^4|\xi|^2+\mu)}.
\eea
Inflation can take place

(i) in the vacuum-dominated region,

(ii) around the local minimum,

(iii)  in the large field inflation.

In the region (i),
$\frac{d{\hat V}}{d\phi}\simeq
-\kappa^2 |\xi|\lambda v^4 \phi <0$,
and $\phi$ rolls down toward $\phi=\phi_{e}$.
To obtain a 
stable potential,
it must take place for $\phi>\phi_c$,
where $\chi$ field has a positive mass.
Then,
inflation is
the type (II).
Around $\phi=0$,
the slow-roll parameters become
\bea
\epsilon
\simeq 8\kappa^2 |\xi|^2 \phi^2, \quad
\eta\simeq -4|\xi|.
\eea
Inflation is terminated
when $\epsilon$ becomes order unity,
namely at
\bea
\phi=\phi_f:= \frac{1}{2\sqrt{2}\kappa |\xi|}.
\eea
The e-folding number is given by
\bea
\label{cons}
N\simeq \frac{1}{4|\xi|}
\simeq \frac{\kappa\phi_f}{\sqrt{2}}.
\eea
Therefore, to get  sufficiently long inflation, $\phi_f\gg \frac{1}{\kappa}$.
On the other hand,
since $\phi_e>\phi_f$, we have to require $\phi_e\gg \frac{1}{\kappa}$, which
leads to
$\frac{|\xi|}{\mu}>\frac{1}{\lambda \kappa^4 v^4}$.
Combining with (\ref{cons}), we obtain
$N<\frac{\lambda\kappa^4  v^4}{4\mu}$.
We have to require $N>50$, and hence
$\frac{\lambda \kappa^4v^4}{\mu}>200$.
Recalling the sub-Planckian condition Eq. (\ref{qg}), unless $\mu\ll 1$,
this condition cannot be satisfied.
The amplitude of the curvature perturbations, the tensor-to-scalar ratio,
and the spectral indices of the curvature and tensor perturbations
are given by
\bea
P_{s}&\simeq &\frac{\kappa^2 \lambda v^4}
    {768 \pi^2 |\xi|^2\phi_{\ast}^2},
\quad
r\simeq 128 \kappa^2 |\xi|^2 \phi_{\ast}^2,\quad
n_{s}-1\simeq  -8 |\xi| < 0, \quad
n_t\simeq -16 \kappa^2 |\xi|^2 \phi_{\ast}^2,
\eea
respectively,
where the field value $\phi_{\ast}$ is evaluated
at the horizon crossing.
The spectrum of the curvature perturbations is red tilted.
In this case, however, since $V_e>0$,
there is no way to terminate inflation
in the original system.

In the region (ii)
where $\phi$ field, located at $\phi > \phi_e$, begins to roll
 toward to $\phi_e$,
inflation can be the type (I).
The tachyonic instability appears
at $\phi=\phi_c$,
and we need to require $\phi_c>\phi_{e}$, or
equivalently
\bea
|\xi|<\frac{\mu}{\kappa^2 v^2 g^2}.
\eea
Taking the sub-Planckian condition Eq. (\ref{qg})
into consideration,
we can further restrict the range of the coupling.
If
$\mu^{\frac{1}{2}}
<\frac{m^2}{\lambda v^4}$,
then the type (I)
inflation occurs for all the allowed region of Eq. (\ref{yamada}).
If
$
\mu^{\frac{1}{2}}
< \frac{\mu}{\kappa^2 v^2 g^2}$,
we obtain
\bea
\mu^{\frac{1}{2}}
<|\xi| <  \frac{\mu}{\kappa^2 v^2 g^2}.
\eea
If
$\mu^{\frac{1}{2}}>\frac{\mu}{\kappa^2 v^2 g^2}$
there is no region where the type (I) inflation can occur at the
sub-Planckian energy scale.

In the region (ii),
assuming $\phi_{e}\lesssim \phi_c$,
the slow-roll parameters
\bea
\epsilon&\simeq& \frac{32\kappa^2 \mu^3|\xi|^2(\phi-\phi_{e})^2}
 {(\mu+|\xi|^2\kappa^4\lambda v^4)^2
 \big(\mu+|\xi|^2\kappa^4\lambda v^4(1+6|\xi|)\big)},
\nonumber\\
\eta&\simeq &\frac{8|\xi|\mu}
   {\mu+|\xi|^2\kappa^4\lambda v^4(1+6|\xi|)}>0.
   \label{eq:slowroll_m0ii}
\eea
Note that as $\phi\to\phi_e$,
$\epsilon$ becomes smaller, and
there is no violation of the slow-roll conditions.
Therefore, inflation must be the type (I).
The e-folding number is given by
\bea
N
&\simeq &
\frac{\kappa^4 |\xi|^2\lambda v^4
\big(1+6|\xi|\big)
+\mu}
{8|\xi|\mu}.
\eea
The amplitude and spectral index of the curvature perturbations
are given by
\bea
P_{s} &\simeq &
\frac{v^4\kappa^2\lambda (|\xi|^2\kappa^4 \lambda^4+\mu)
(\mu+\kappa^2 \lambda v^4|\xi|^2(1+6|\xi|))}
{3072|\xi|^2\pi^2\mu^2(\phi_\ast-\phi_{e})^2},
\nonumber\\
n_{s}-1&\simeq&  \frac{16|\xi|\mu}
     {\mu+\kappa^4 \lambda v^4|\xi|^2(1+6|\xi|)} \simeq \frac{2}{N},
\nonumber \\
n_t &\simeq&
-\frac{64\kappa^2 \mu^3|\xi|^2(\phi_\ast-\phi_{e})^2}
 {(\mu+|\xi|^2\kappa^4\lambda v^4)^2
 \big(\mu+|\xi|^2\kappa^4\lambda v^4(1+6|\xi|)\big)} 
\simeq -2\epsilon,
\nonumber \\
r &\simeq& 16\epsilon.
\eea
The spectral index of the curvature perturbations
is blue tilted.

In the region (iii),
$\phi\gg \phi_{e}$,
the slow-roll parameters Eq. (\ref{sl}) are given by
\bea
&&\epsilon
\simeq \frac{8}
            {(1+6|\xi|)|\xi|\kappa^4 \phi^4},
\quad
\eta
\simeq -\frac{8}{(1+6|\xi|)\kappa^2 \phi^2} < 0,
\eea
and the slow-roll approximation is valid in the range
\bea
\phi>\phi_f=\frac{2\sqrt{2}}
{(6|\xi|+1)^{\frac{1}{2}}\kappa}.
\eea
Inflation can be
either the type  (I) or the type (II).
If $\phi_c>\phi_f$, inflation is the type (II),
while
if $\phi_c<\phi_f$ it is the type (I).
The e-folding number
from the horizon crossing to the end of inflation
is given by
\bea
N\simeq \frac{(1+6|\xi|)\kappa^2\mu}
       {8\mu}\Big(\phi_\ast^2-\phi_{c,f}^2\Big)
\simeq
\frac{(1+6|\xi|)\kappa^2\phi_{\ast}^2}
       {8},
\eea
where we assume that $\phi_{\ast}\gg\phi_{c,f}$.
The amplitude of the curvature perturbations
and the tensor-to-scalar ratio are given by
\bea
&&P_{s}
\simeq \frac{(1+6|\xi|)\kappa^4 \mu\phi_{\ast}^4}
   {768|\xi|\pi^2},
\quad
r\simeq\frac{128}
       {|\xi| (1+6|\xi|)\kappa^4\phi_{\ast}^4}
\simeq \frac{2}{N^2}\big(6+\frac{1}{|\xi|}\big).
\eea
The spectral indices of the curvature and tensor perturbations become
\bea
&&n_{s}-1
\simeq -\frac{16}
             {(1+6|\xi|)\kappa^2\phi_{\ast}^2}
\simeq  -\frac{2}{N}<0,
\nonumber\\
&&n_t\simeq
-\frac{16}
            {(1+6|\xi|)|\xi|\kappa^4\phi_{\ast}^4}
\simeq
-\frac{1+6|\xi|}{4|\xi|N^2}<0.
\eea
Thus, the spectrum of the curvature perturbations
is red tilted.

We perform the numerical calculations to track the evolution of $\phi$
field along $\chi = 0$ and  of slow-roll parameters $\epsilon$ and $\eta$
in Fig. \ref{figm0-1}. 
We set $\lambda = g^2 = 1, v^2 = 10^{-4} m_{pl}^2$ and $\mu = 10^{-8}$
in Fig. \ref{figm0-1}.
We find that the local minimum locates at $\phi = \phi_e$
in Fig. \ref{figm0-1}-(a). 
 Given parameters for Fig. \ref{figm0-1}, as seen (\ref{eq:slowroll_m0ii}),
 slow-roll conditions are not
violated (see Fig. \ref{figm0-1}-(b) and \ref{figm0-1}-(c)),
 so inflation should 
be terminated by the tachyonic instability.
In order to terminate inflation by the tachyonic instability,
it is necessary to satisfy the condition
$\phi_c > \phi_e$.
This condition leads to
\bea
|\xi| < |\xi_c| \equiv \frac{\mu}{\kappa^2 v^2 g^2}.
\label{eq:xic}
\eea
$|\xi_c| \sim 4\times 10^{-6}$ for the Fig. \ref{figm0-1}.
In Fig. \ref{figm0-1}-(d) 
we plot $n_s -r$ relations. The figure shows that spectrum changes
from red spectrum to blue ones.

\subsection{$\phi^{2p}$ ($p>2$) model}

Before closing this subsection, we briefly discuss
the case of the potential in the Jordan frame
given by
\bea
V(\phi,\chi)=
\frac{1}{4}\lambda \big(\chi^2-v^2\big)^2
+\frac{1}{2p}\mu \phi^{2p}
+\frac{1}{2}g^2\phi^2\chi^2,
\eea
where $p>2$ is an integer.
Note that in this subsection $\mu$
parameter has mass dimension of $(-2(p-2))$,
and in particular $(-4)$ for $p=3$.
In this case,
the potential in the Einstein frame $\hat V$ has a
minimum at some $\phi=\phi_{e}>0$.
For $p=3$,  if $-1<\frac{A}{2\mu}<1$, where
\bea
A:= 3|\xi|^2v^4\kappa^6\lambda-2\mu,
\eea
the extremal point is given by
\bea
\phi_e:=\frac{1}{\kappa|\xi|^{\frac{1}{2}}}
\Big(-1+2\cos\big(\frac{x}{3}\big)\Big)^{\frac{1}{2}},
\eea
where $\cos x:=\frac{A}{2\mu}$.
Note that $\frac{1}{2}<\cos\big(\frac{x}{3}\big)<1$.
On the other hand, if $\frac{A}{2\mu}>1$, then
it is given by
\bea
\phi_e:=\frac{1}{\kappa|\xi|^{\frac{1}{2}}}
\Big(-1+2\cosh\big(\frac{y}{3}\big)\Big)^{\frac{1}{2}},
\eea
where $\cosh y:=\frac{A}{2\mu}$.


In the vacuum-dominated region,
${\hat V}$ is always negatively tilted. 
And thus inflation is the type (II)
and is terminated at
\bea
\phi=\phi_f:=\frac{1}{2^{\frac{3}{2}}|\xi|\kappa}.
\eea
The typical e-folding number is given by
\bea
N\simeq \frac{1}{4|\xi|}
\simeq\sqrt{2} \kappa \phi_f.
\eea
To realize sufficiently long inflation, we require
$\phi_f>\frac{1}{\kappa}$, 
which leads to $|\xi|< 1$ for $\frac{A}{2\mu}<1$
and $|\xi|<\frac{v^4\kappa^6\lambda}{\mu}$ for $\frac{A}{2\mu}>1$.
Thus, we obtain $N> 1$ for $\frac{A}{2\mu}<1$
and $N> \frac{\mu}{v^4\kappa^6\lambda}$
for $\frac{A}{2\mu}>1$.


Around the local minimum,
it is naturally expected that
the spectrum of the curvature perturbations
produced inflation near $\phi=\phi_{e}$
is always blue tilted.

In the large field region,
since $\frac{d{\hat V}}{d{\hat\phi}}>0$,
$\phi$ is decreasing.
The slow-roll parameters are given by
\bea
\epsilon&\simeq&
\frac{2(p-2)^2|\xi|}{1+6|\xi|}
+\frac{2(p-2)\big((p+2)+12p|\xi|\big)}
{\kappa^2 (1+6|\xi|)^2 \phi^2},
\nonumber\\
\eta&\simeq&
\frac{4(p-2)^2|\xi|}{1+6|\xi|}
+\frac{2(2p-5)\big((p+2)+12p|\xi|\big)}
{\kappa^2 (1+6|\xi|)^2 \phi^2}.
\eea
The slow-roll conditions is violated when $\epsilon$
or $\eta$ becomes $O(1)$
at
\bea
\phi\simeq
\phi_f:=\frac{\sqrt{2(p-2)\big((p+2)+12p|\xi|\big)}}
                  {(1+6|\xi|)\kappa}.
\eea
For $\phi_c>\phi_f$, inflation is the type (I),
and
for $\phi_c<\phi_f$, it is the type (II).
The e-folding number is given by
\bea
N
\simeq \frac{1+6|\xi|}{2|\xi|(p-2)}.
\eea
The spectrum of the curvature perturbations,
the tensor-to-scalar ratio,
the spectral indices of the curvature and tensor
perturbations are given by
\bea
&&
P_{s}\simeq \frac{(1+6|\xi|)\mu\phi_{\ast}^{2(p-2)}}
{96|\xi|^3(p-2)^2p\pi^2},
\quad
r\simeq \frac{32|\xi|(p-2)^2}{1+6|\xi|},
\nonumber\\
&&
n_{s}-1\simeq -\frac{4(p-2)^2|\xi|}{1+6|\xi|}
\simeq -\frac{2(p-2)}{N},
\quad
n_t\simeq -\frac{4(p-2)^2|\xi|}{1+6|\xi|}
\simeq -\frac{2(p-2)}{N}.
\eea
Thus, the spectrum of the curvature perturbations is red tilted.


\section{The case of $m \neq 0, \mu \neq 0$}

In this section, we discuss inflationary dynamics
in the Einstein frame in more details.
Along $\chi=0$,
the potential in the Einstein frame has the nonzero
extremal value
\bea
{\hat V}(\phi_e,0)=
{\hat V}_e
:=\frac{m^4-\mu\lambda v^4}
{4\big[|\xi|\kappa^2 m^2-\mu
+|\xi|\kappa^2 (m^2-|\xi|\kappa^2\lambda v^4)\big]},
\eea
at
\bea
\phi=
\phi_{e}:=\sqrt{\frac{-m^2-\kappa^2\lambda v^4 \xi}{\mu+\kappa^2m^2\xi}}.
\label{past}
\eea
For $\frac{m^4}{\mu \lambda v^4}<1$,
$\hat V_e$ becomes the minimum,
while for $\frac{m^4}{\mu \lambda v^4}>1$,
it becomes the maximum.
For $\frac{m^4}{\mu\lambda v^4}=1$,
the potential has no extremal point.

\subsection{$\frac{m^4}{\mu\lambda v^4}<1$}
\label{sect:a}

\begin{figure}
\centering
\includegraphics[width=0.8\textwidth]{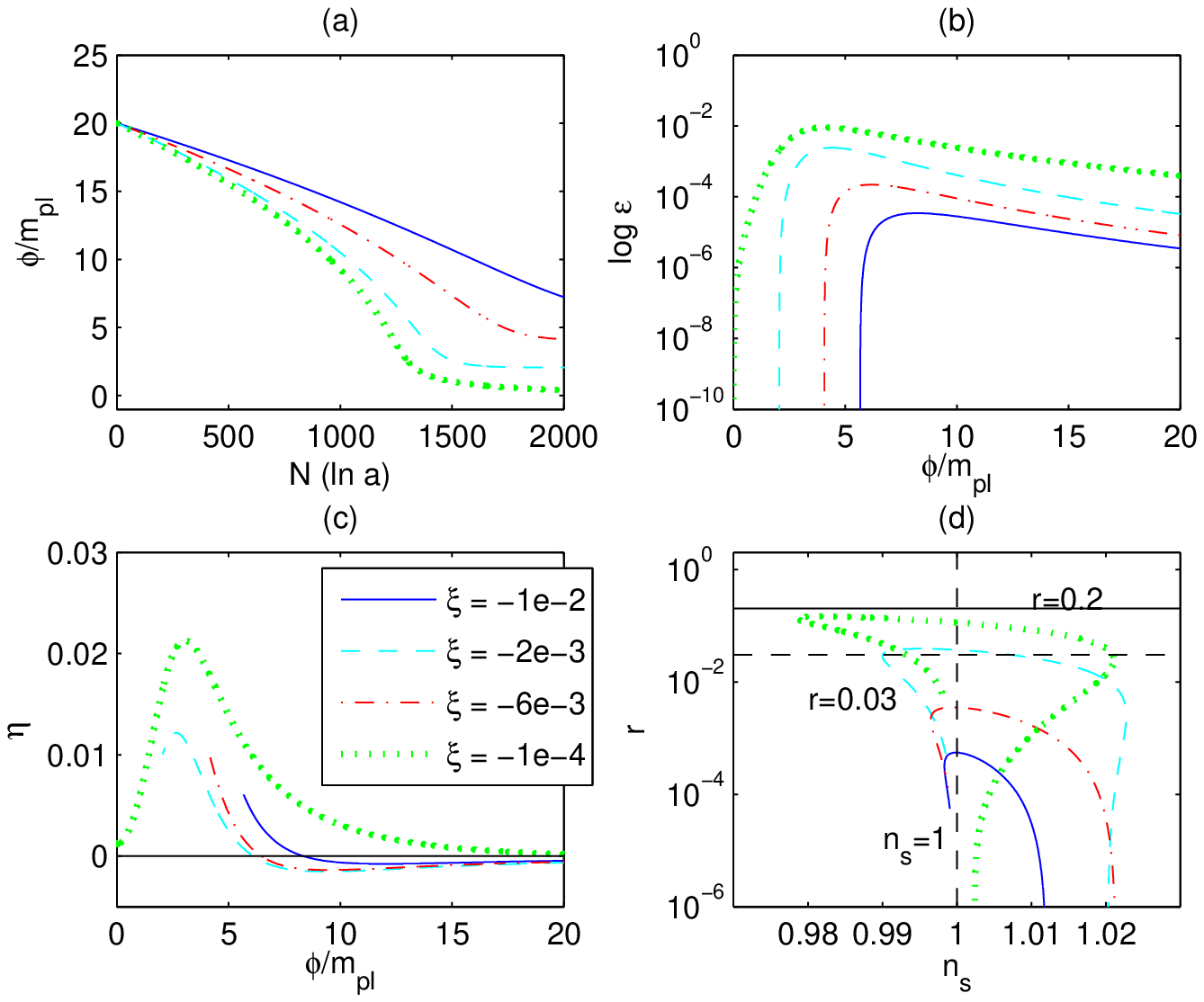}
\caption{ We plot (a) $\phi$ field evolution with the number of efolds $N$,
(b) evolution of slow roll parameter $\epsilon$ with $\phi$,
(c) evolution of slow-roll parameter $\eta$ with $\phi$, and (d)
spectral index of scalar perturbations $n_s$ versus tensor to scalar ratio $r$
with $\xi = -10^{-2}, -2\times 10^{-3}, -6\times 10^{-3}$ and $-10^{-4}$.
We set $\lambda = g^2 =1,\,\, v^2 = 10^{-2}m_{pl}^2,\,\, m^2 = 10^{-6}m_{pl}^2,
\,\, \mu = 10^{-6}$. These parameter satisfy the condition
$m^4 < \mu\lambda v^4$ discussed in Sect. \ref{sect:a}.
While the potential is monotonically increasing  for $\xi
= -10^{-4}$,  the potential has one local minimum at $\phi = \phi_e$
for $\xi = -10^{-2}, -2\times 10^{-3}$ and $-6\times 10^{-3}$. 
As $|\xi|$ decreases,
the local minimum shifts to the smaller $\phi$ and if $|\xi| \leq
3\times 10^{-4}$ the potential behaves as a monotonically
increasing function of $\phi$. }
\label{fig1}
\end{figure}

\begin{figure}
\centering
\includegraphics[width=0.8\textwidth]{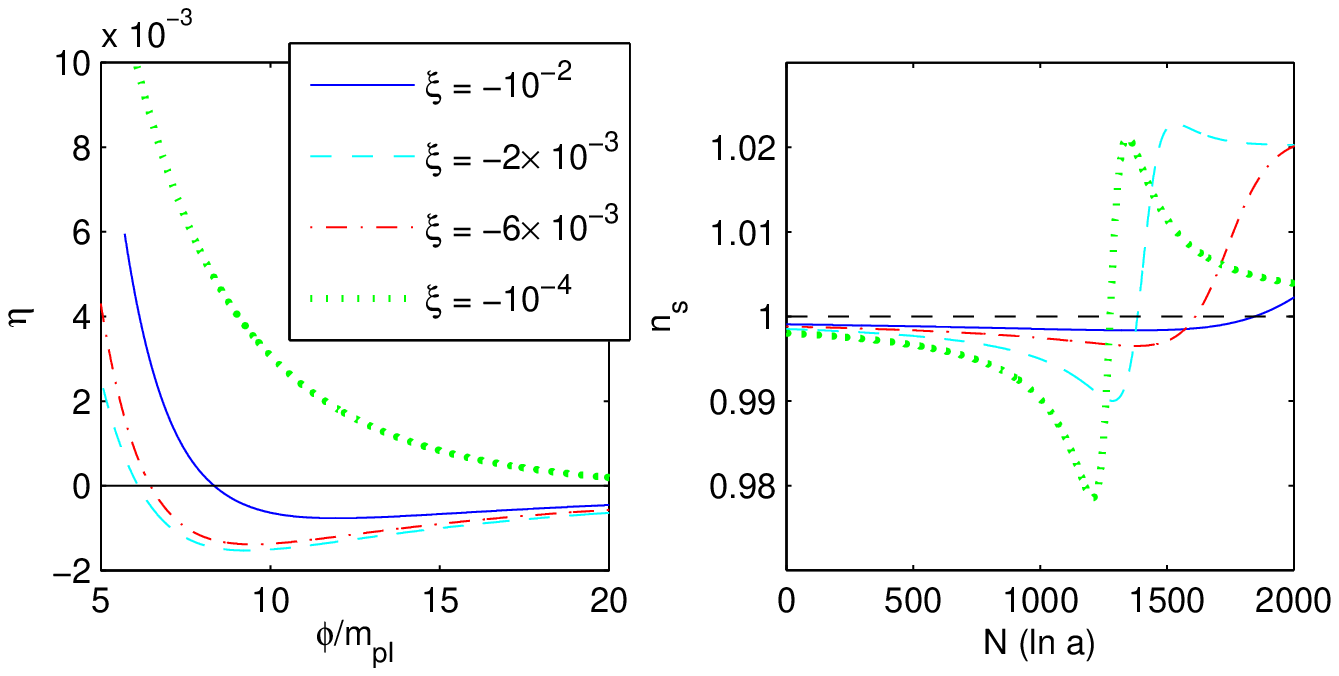}
\caption{ (left) Magnifying the Fig. \ref{fig1}-(c) in the region
$5<\phi < 20$. (right) evolution of spectral index $n_s$
of scalar perturbation.}
\label{fig2}
\end{figure}

\begin{figure}
\centering
\includegraphics[width=0.8\textwidth]{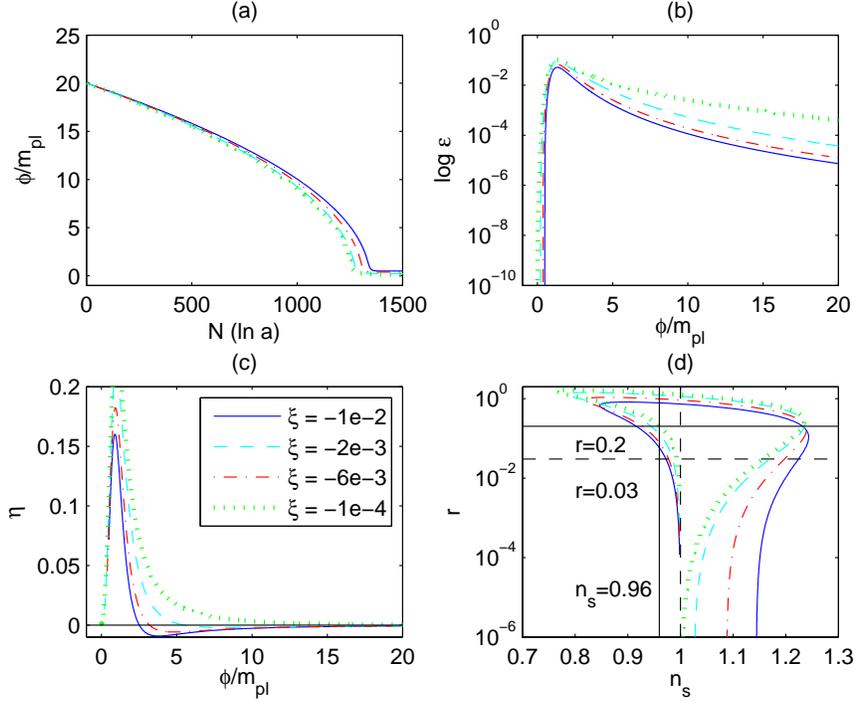}
\caption{Same as Fig. \ref{fig1} but we set
$\lambda = g^2 =1,\,\, v^2 = 10^{-2} \, m_{pl}^2,\,\, m^2 = 10^{-6}\, m_{pl}^2,
\,\,\mu = 10^{-4}$ which satisfy $m^4 < \mu\lambda v^4$ (see
 Sect. \ref{sect:a}) .
For $\xi = -10^{-4}$, the potential is monotonically increasing.
 For $\xi = -10^{-2}, -2\times 10^{-3}$ and $-6\times 10^{-3}$, 
the potential has
one local minimum at $\phi = \phi_e$.  }
\label{fig3}
\end{figure}
\begin{figure}
\centering
\includegraphics[width=0.8\textwidth]{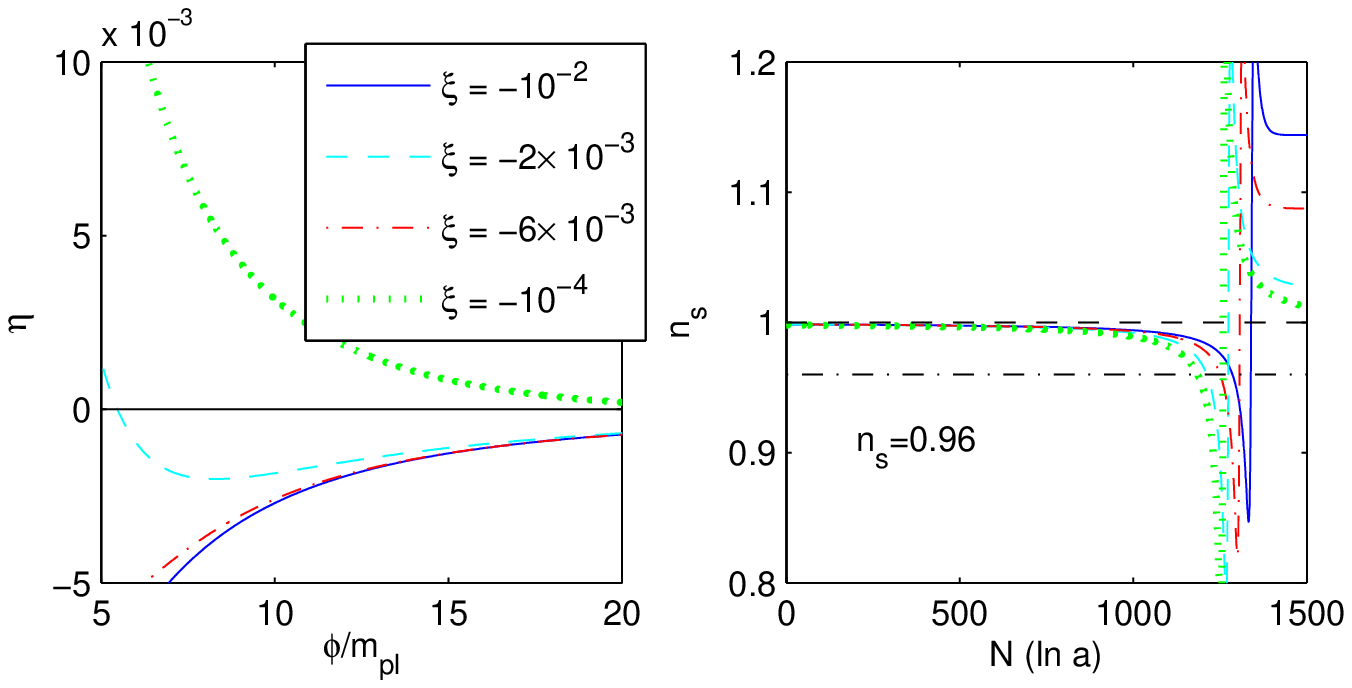}
\caption{ (left) Magnifying the Fig. \ref{fig3}-(c) in the region
$5<\phi < 20$. (right) evolution of spectral index $n_s$
 of scalar perturbation.}
\label{fig4}
\end{figure}


In this case,
in terms of the shape of $\hat V$,
we separately discuss the following cases

(1) if $\frac{m^2}{\kappa^2 \lambda v^4}<|\xi|<\frac{\mu}{\kappa^2 m^2}$,
there is one local minimum in the potential at $\phi=\phi_e$:

(2) if $\frac{m^2}{\kappa^2 \lambda v^4}>|\xi|>0$,
the potential is monotonically increasing:

(3) if $|\xi|> \frac{\mu}{\kappa^2m^2}$,
the potential is monotonically decreasing:

\subsubsection*{\rm (1):}

In this case, there is the potential minimum at
$\phi=\phi_e$.
In the typical cases,
inflation can take place

(i) in the vacuum-dominated region,

(ii) around the local minimum,

(iii)  in the large field inflation.

In the region (i),
$\frac{d{\hat V}}{d\phi}\simeq
(m^2-\kappa^2 |\xi|\lambda v^4) \phi <0$,
and $\phi$ rolls down toward $\phi=\phi_{e}$.
To obtain a stable potential,
it must take place for $\phi>\phi_c$.
Then,
inflation is 
the type (II).
Around $\phi=0$,
the slow-roll parameters become
\bea
\epsilon\simeq \frac{8(|\xi|\kappa^2\lambda v^4-m^2)^2\phi^2}
                     {v^8\kappa^2 \lambda^2},\quad
\eta\simeq -4(|\xi|-\frac{m^2}{v^4\kappa^2\lambda}).
\eea
Inflation is terminated
where $\epsilon$ becomes order unity,
namely at
\bea
\phi=\phi_f:= \frac{\kappa\lambda v^4}
{2\sqrt{2}(|\xi|\kappa^2\lambda v^4-m^2)}.
\eea
The e-folding number is given by
\bea
N\simeq \frac{\kappa^2\lambda v^4}
       {4(\kappa^2 \lambda |\xi|v^4-m^2)}
\simeq \frac{\kappa\phi_f}{\sqrt{2}}.
\label{eq:efold_a1}
\eea
Therefore, to get sufficiently long inflation, $\phi_f\gg \frac{1}{\kappa}$.
On the other hand, we also require
$\phi_e>\phi_f$, hence $\phi_e\gg \frac{1}{\kappa}$, which leads to
$$
\frac{\mu}{\kappa^2 m^2}> |\xi|>\frac{\kappa^2 m^2+\mu}
{\kappa^2 m^2+ \kappa^2 \lambda v^4},
$$
where the upper bound given from our classification.
Combining with (\ref{eq:efold_a1}) 
we obtain
$$
N< \frac{\kappa^2 m^2+ \kappa^4 \lambda v^4}
     {4(\mu-\frac{m^4}{\lambda v^4})}.
$$
We have to require $N>50$, hence
$$
\frac{\kappa^2 m^2 + \kappa^4\lambda v^4}
     {\mu-\frac{m^4}{\lambda v^4}}>200.
$$
For $\mu\gtrsim \frac{m^4}{\lambda v^4}$,
this condition can be satisfied.
In the limit of $m\to 0$,
we recover the result in Sec. III-B.

The amplitude of the curvature perturbations, the tensor-to-scalar ratio,
and the spectral indices of the curvature and tensor perturbations
are given by
\bea
P_{s}&\simeq &\frac{v^{12}\kappa^6 \lambda^3}
    {768 \pi^2\big(m^2-|\xi|\kappa^2\lambda v^4\big)^2\phi_{\ast}^2},
\quad
r\simeq \frac{128(m^2-|\xi|\kappa^2 \lambda v^4)^2\phi_\ast^2}
       {v^8\kappa^2\lambda^2},
\nonumber\\
n_{s}-1&\simeq &-8\big(|\xi|-\frac{m^2}{\kappa^2 \lambda v^4}\big)<0,\quad
n_t \simeq -\frac{16(m^2-|\xi|\kappa^2 \lambda v^4)^2\phi_{\ast}^2}
          {\kappa^2\lambda^2 v^8}\,,
\eea
respectively,
where the field value $\phi_{\ast}$ is evaluated
at the horizon crossing.
Thus, the spectrum of the curvature perturbations is red tilted.

In the two regions of (ii) and (iii),
inflation can be the type (I).
Let us clarify the conditions that
the type (I) inflation takes place.
The tachyonic instability appears
at the place where $\chi$ deviates from $\chi=0$.
Then, the critical field value must be greater than $\phi_e$,
hence $\phi_c>\phi_{e}$, which leads to
\bea
|\xi|<  |\xi_c| :=
\frac{g^2m^2+\lambda\mu v^2}{\kappa^2\lambda v^2(m^2+g^2 v^2)}.
\eea
This is the generalization of Eq. (\ref{eq:xic}) 
to the case of $m \neq 0$.
Noting
\bea
\frac{\mu}{m^2}
> |\xi_c|
>\frac{m^2}{\lambda v^4},
\eea
the type (I) inflation can take place for
\bea
\frac{m^2}{\kappa^2 \lambda v^4}
<|\xi|
< |\xi_c|.
\label{yamada}
\eea

Taking the sub-Planckian condition Eq. (\ref{qg})
into consideration,
we can further restrict the range of the coupling.
If
$\mu^{\frac{1}{2}}
<\frac{m^2}{\lambda v^4}$,
then the type (I)
inflation occurs for all the allowed region of Eq. (\ref{yamada}).
If
$
\frac{m^2}{\kappa^2\lambda v^4}
<\mu^{\frac{1}{2}} < |\xi_c|$,
we obtain
\bea
\mu^{\frac{1}{2}}
<|\xi| < |\xi_c|.
\eea
If
$\mu^{\frac{1}{2}}> |\xi_c|$,
there is no region where the type (I) inflation can occur at the
sub-Planckian energy scale.

In the region (ii),
assuming $\phi_{e}\lesssim \phi_c$,
the slow-roll parameters
\bea
\epsilon&\simeq&
32(|\xi|\kappa^2 v^4\lambda-m^2)^2(\mu-|\xi|\kappa^2 m^2)^5
(\phi-\phi_{e})^2
\nonumber\\
&\times &
\Big\{\kappa^2
(\kappa^2m^2|\xi|-\mu
+|\xi|\kappa^2(m^2-|\xi|\kappa^2v^4))^2
\nonumber\\
&\times&
\big[
(|\xi|\kappa^2 v^4\lambda-m^2)(1+6|\xi|)\kappa^2|\xi|
+\mu-m^2\kappa^2|\xi|\big]
(\lambda v^4\mu-m^4)^2
\Big\}^{-1}
\nonumber\\
\eta
&\simeq&
\frac{8(|\xi|\kappa^2 \lambda v^4-m^2)
       (\mu-|\xi|\kappa^2 m^2)^2}
 {\kappa^2
\big[
(|\xi|\kappa^2 v^4\lambda-m^2)(1+6|\xi|)\kappa^2|\xi|
+\mu -m^2\kappa^2|\xi|
\big](\lambda v^4\mu-m^4)}
>0.
\eea
Note that as $\phi\to\phi_e$,
$\epsilon$ becomes smaller, and
there is no violation of the slow-roll conditions.
Therefore, inflation must be the type (I).
The e-folding number is given by
\bea
N
&\simeq &
\frac
{\kappa^2
\big[
(|\xi|\kappa^2 v^4-m^2\lambda)(1+6|\xi|)\kappa^2|\xi|
+\mu-m^2\kappa^2|\xi|
\big](\lambda v^4\mu-m^4)}
{8(|\xi|\kappa^2 \lambda v^4-m^2)
       (\mu-|\xi|\kappa^2 m^2)^2}.
\eea
The amplitude and spectral index of the curvature perturbations
are given by
\bea
P_{s}
&=&\kappa^6
\Big[
(-2|\xi|\kappa^2 m^2+|\xi|^2\kappa^4\lambda v^4+\mu)
\big(-2|\xi|\kappa^2 m^2+6|\xi|^3 v^4\kappa^4\lambda
+\kappa^2 |\xi|(\kappa^2\lambda v^4-6m^2 )\big)
\nonumber\\
&\times&
(m^4-\lambda \mu v^4)^3
\Big]
\nonumber\\
&\times&
\Big[
3072\pi^2
(m^2-|\xi|\kappa^2\lambda v^4 )^2
(|\xi\kappa^2m^2-\mu)^5
(\phi_{\ast}-\phi_e)^2
\Big]^{-1},
\nonumber\\
n_{s}-1&\simeq&
\frac{16(|\xi|\kappa^2 \lambda v^4-m^2)(\mu-|\xi|\kappa^2 m^2)^2}
 {\kappa^2
\big[
(|\xi|\kappa^2 v^4\lambda-m^2)(1+6|\xi|)\kappa^2|\xi|
+\mu-m^2\kappa^2|\xi|
\big](\lambda v^4\mu-m^4)}
\simeq \frac{2}{N}>0.
\eea
The spectral index of the curvature perturbations
is blue tilted.
The tensor spectral index and the tensor-to-scalar ratio
are given by
\bea
n_{T}\simeq  -2\epsilon,\quad
r\simeq 16\epsilon.
\eea

In the region (iii),
$\phi\gg \phi_{e}$,
the slow-roll parameters Eq. (\ref{sl}) are given by
\bea
&&\epsilon
\simeq \frac{8\big(\mu-|\xi|\kappa^2m^2\big)^2}
            {(1+6|\xi|)|\xi|\kappa^4 \mu^2\phi^4},
\quad
\eta
\simeq -\frac{8\big(\mu-|\xi|\kappa^2m^2\big)}
            {(1+6|\xi|)\kappa^2 \mu \phi^2},
\eea
and the slow-roll approximation is valid in the range
\bea
\phi>\phi_f=\frac{2^{3/2}(\mu-|\xi|\kappa^2 m^2)^{1/2}}
{(6|\xi|+1)^{1/2}(\kappa^2\mu)^{1/2}}.
\eea
Inflation can be
either the type  (I) or the type (I).
If $\phi_c>\phi_f$, inflation is the type (II),
while
if $\phi_c<\phi_f$ it is the type (I).
The e-folding number
from the horizon crossing to the end of inflation
is given by
\bea
N\simeq \frac{(1+6|\xi|)\kappa^2\mu}
       {8(\mu-|\xi|\kappa^2 m^2)}\Big(\phi_\ast^2-\phi_{c,f}^2\Big)
\simeq
\frac{(1+6|\xi|)\kappa^2\mu\phi_{\ast}^2}
       {8(\mu-|\xi|\kappa^2 m^2)},
\eea
where we assume that $\phi_{\ast}\gg\phi_{c,f}$.
The amplitude of the curvature perturbations
and the tensor-to-scalar ratio are given by
\bea
&&P_{s}
\simeq \frac{(1+6|\xi|)\kappa^4 \mu^3\phi_{\ast}^4}
   {768|\xi|\pi^2\big(\mu-|\xi|\kappa^2 m^2\big)^2},
\quad
r\simeq\frac{128(\mu-|\xi|\kappa^2m^2)^2}
       {|\xi| (1+6|\xi|)\kappa^4\mu^2\phi_{\ast}^4}
\simeq \frac{2}{N^2}\big(6+\frac{1}{|\xi|}\big).
\eea
The spectral indices of the curvature and tensor perturbations become
\bea
&&n_{s}-1
\simeq -\frac{16\big(\mu-|\xi|\kappa^2m^2\big)}
             {(1+6|\xi|)\kappa^2 \mu\phi_{\ast}^2}
\simeq  -\frac{2}{N}<0,
\nonumber\\
&&n_t\simeq
-\frac{16\big(\mu-|\xi|\kappa^2m^2\big)^2}
            {(1+6|\xi|)|\xi|\kappa^4 \mu^2\phi_{\ast}^4}
\simeq
-\frac{1+6|\xi|}{4|\xi|N^2}<0.
\eea
Thus, the spectrum of the curvature perturbations
is red tilted.

\subsubsection*{\rm (2):}
The potential is monotonically increasing as $\phi$ increases,
since $\phi_e \to 0$,
as $\xi\to -\frac{m^2}{\kappa^2 \lambda v^4}$ in the case (1).
In this case, inflation can take place

(i) in the vacuum-dominated region

(ii) in the large field region

In the region (i),
$\frac{d{\hat V}}{d\phi}\simeq
(m^2-\kappa^2 |\xi|\lambda v^4) \phi >0$,
and $\phi$ slowly rolls down 
toward the origin.
Then, the slow-roll parameters
are given by
\bea
\epsilon
\simeq \frac{8(m^2-|\xi|\kappa^2 \lambda v^4)^2\phi^2}
              {\kappa^2 \lambda^2 v^8},
\quad
\eta\simeq
\frac{4(m^2-|\xi|\kappa^2 \lambda v^4)}
              {\kappa^2 \lambda v^4}>0.
\eea
Since $\epsilon$ is decreasing for decreasing $\phi$,
inflation can be the type (I).
The e-folding number is given by
\bea
N\simeq \frac{\kappa^2\lambda v^4}
       {4(m^2-\kappa^2 \lambda |\xi|v^4)}.
\eea
The slow-roll inflation can be realized
for
\bea
0<\phi<\phi_i
:=\frac{\kappa\lambda v^4}{2\sqrt{2}(-|\xi|\kappa^2\lambda v^4+m^2)},
\eea
where $\epsilon_i=1$.
Noting that $N\simeq \frac{\kappa \phi_i}{\sqrt{2}}$,
to realize sufficiently long inflation,
$\phi_i\gg \frac{1}{\kappa}$, which gives
$$
\frac{m^2 }{\kappa^2\lambda v^4}
>|\xi|>
\frac{m^2 }{\kappa^2\lambda v^4}
-\frac{1}{2\sqrt{2}}
$$
(if $\frac{m^2 }{\kappa^2\lambda v^4}
>\frac{1}{2\sqrt{2}}$),
where the upper bound is given from our classifications.

The amplitude of the curvature perturbations, the tensor-to-scalar ratio,
and the spectral indices of the curvature and tensor perturbations
 are given by
\bea
P_{s}&\simeq &\frac{v^{12}\kappa^6 \lambda^3}
    {768 \pi^2\big(m^2-|\xi|\kappa^2\lambda v^4\big)^2\phi_\ast^2},
\quad
r\simeq \frac{128(m^2-|\xi|\kappa^2 \lambda v^4)^2\phi_\ast^2}
       {v^8\kappa^2\lambda^2},
\nonumber\\
n_{s}-1&\simeq &8\big(\frac{m^2}{\kappa^2 \lambda v^4}-|\xi|\big)>0,\quad
n_t\simeq -\frac{16(m^2-|\xi|\kappa^2 \lambda v^4)^2}
          {\kappa^2\lambda^2 v^8}\,.
\eea
Thus, the spectrum of the curvature perturbations is blue tilted.

In the region (ii),
the background dynamics and predictions can be written
as the same manner as those
in (1)-(iii).
We now discuss the condition that inflation takes place
at the sub-Planckian scale.
If $\mu^{\frac{1}{2}}<\frac{m^2}{\kappa^2\lambda v^4}$,
then inflation takes place
at a sub-Planckian energy scale
for
\bea
\mu^{\frac{1}{2}}<|\xi|<
\frac{m^2}{\kappa^2 \lambda v^4}.
\eea
For
$\mu^{\frac{1}{2}}>\frac{m^2}{\kappa^2\lambda v^4}$,
there is no region where inflation can occur in the
sub-Planckian energy scale.

Until now we have discussed inflation realized in each typical region.
In these cases, in general $\phi$ field moves
from the large field region
toward the local minimum or to the vacuum dominated region.
To follow the whole dynamics,
we need to solve the equations of motion numerically.

In Fig. \ref{fig1} and \ref{fig3}, we plot $\phi$ field
evolution with the number of e-folds $N$ in panel (a),
evolution of slow-roll parameters $\epsilon$ and $\eta$ in
panel (b) and (c), and the relation between 
the spectral index of scalar perturbations
$n_s$  and the tensor to scalar ratio $r$ in panel (d)
with $\xi = -10^{-2}, -2\times 10^{-3}, -6\times 10^{-3}$ and $-10^{-4}$.
We set the parameters to $\lambda = 1,\,\, g^2 = 1,\,\,
 v^2 = 10^{-2}\, m_{pl}^2,\,\,
m^2 = 10^{-6}\, m_{pl}^2,\,\, \mu = 10^{-6}$ for Fig. \ref{fig1}
and to $\lambda = 1,\,\, g^2 = 1,\,\, v^2 = 10^{-2}\, m_{pl}^2,\,\,
m^2 = 10^{-6} \,m_{pl}^2,\,\, \mu = 10^{-4}$ for Fig. \ref{fig3}. Both
parameter set satisfy the condition $m^4 < \mu \lambda v^4$.
In Fig. \ref{fig1}, 
we find that as $|\xi|$
decreases, the local minimum value $\phi_e$
 shifts to smaller $\phi$. For $\xi =
-10^{-4}$, the potential increases monotonically ($\phi_e  = 0$).
If $|\xi| < \frac{m^2}{\kappa^2 \lambda v^4} \simeq
4 \times 10^{-4}$, the potential behaves as an increasing
function of $\phi$. If $|\xi| \geq \frac{m^2}{\kappa^2 \lambda v^4} \simeq
4 \times 10^{-4}$, there exists a local maximum
 at $\phi = \phi_e$. On the other hand,
if $|\xi| \geq \frac{\mu}{\kappa^2 m^2} \simeq 4\times 10^{-2}$,
the potential decreases monotonically as $\phi$ increases which will
discuss below.
We find in Fig. \ref{fig1}-(a)
 the $\phi$  field reach to $\phi_e$ after slow-rolling over
potential  during inflation period.

Fig. \ref{fig1}-(b) and \ref{fig1}-(c) show the evolution of the
slow-roll parameters $\epsilon$ and $\eta$. These figures show that
slow-roll conditions are not violated for the chosen parameter, 
so inflation should be 
terminated by the tachyonic instability.
 We find in Fig. \ref{fig1}-(c) that
the slow-roll parameter $\eta$ has positive values for $\xi = -10^{-4}$
and have negative values for $\xi = -10^{-2}, 2\times 10^{-3}$
and $6\times 10^{-3}$ 
during slow-roll phase. We plot the slow-roll parameter
$\eta$ in detail during slow-roll phase in Fig. \ref{fig2}-(left).
These imply that
the potential have positive curvature ($\xi = -10^{-4}$) or
negative curvature ($\xi = -10^{-2}, 2\times 10^{-3}, 6\times 10^{-3}$)
during inflation.

Fig. \ref{fig1}-(d) shows the relation between the spectral index $n_s$
and the tensor to scalar ratio $r$. In order to compare with the
observations, we draw the bound $r=0.2$ (horizontal black solid line)
from WMAP satellite and $r=0.03$ (horizontal black dashed line) from
PLANCK satellite \cite{Efstathiou:2009xv}. 
Since $n_s = 0.968 \pm 0.012$ from WMAP data \cite{Komatsu:2010fb},
Fig. \ref{fig1}-(d) indicates that $n_s$ is well fit to the
data in $2\sigma\, (95\% CL)$.  We find from Fig. \ref{fig2}-(right)
that the spectral index shows the red spectrum in the large field region
and then moves to the blue spectrum around the local minimum.

In Fig. \ref{fig3}, the potential shows similar behavior
as in Fig. (\ref{fig1}) except that for $|\xi| \geq \frac{\mu}{\kappa^2 m^2}
\simeq 4$
the potential decreases monotonically as $\phi$ increases.
Fig. \ref{fig3}-(c) shows the evolution of the slow-roll parameter $\eta$
and we can look at more closely the behavior of $\eta$ during slow-roll
phase in Fig. \ref{fig4}-(left).

The relation between the spectral index $n_s$
and the tensor to scalar ratio $r$ are shown in Fig. \ref{fig3}-(d)
and the evolution of $n_s$ are plotted in Fig. \ref{fig4}-(right).
The best fit of $n_s$ from WMAP ($n_s = 0.96$) as well as $r =0.2$
and $r=0.03$ are drawn in Fig. \ref{fig3}-(d). Like as in Fig. \ref{fig1},
the figure shows the spectrum moves from red spectrum to blue ones.

\subsubsection*{\rm (3):}
The potential is monotonically decreasing as $\phi$ becomes larger,
since $\phi_{e}\to \infty$,
as $\xi\to -\frac{\mu}{\kappa^2 m^2}$ in the subcase (1).
In this case,
there is no way to realize the reheating,
since the tachyonic instability does not occur.
Thus, in this paper we do not consider this case.

\subsection{ $\frac{m^4}{\mu\lambda v^4}>1$}
\label{sect:b}

\begin{figure}
\centering
\includegraphics[width=0.8\textwidth]{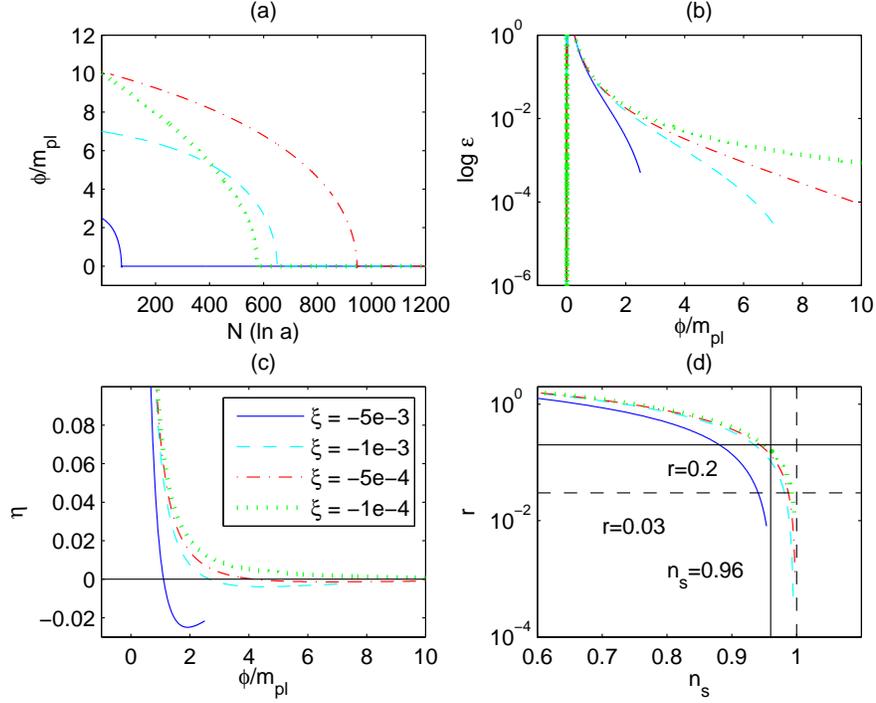}
\caption{We plot (a) $\phi$ field evolution with the number of efolds, $N$,
(b) evolution of slow roll parameter $\epsilon$ with $\phi$,
(c) evolution of slow-roll parameter $\eta$ with $\phi$, and (d)
spectral index of scalar perturbations $n_s$ versus tensor to scalar ratio $r$
with $\xi = -5\times 10^{-3}, -10^{-3}, -5\times 10^{-4}$ and $-10^{-4}$.
We set $\lambda = g^2 =1, v^2 = 10^{-2}m_{pl}^2, m^2 = 10^{-2}m_{pl}^2,
\mu = 10^{-4}$. These parameter satisfy the condition
$m^4 > \mu\lambda v^4$ discussed in Sect. \ref{sect:b}.
For $\xi =  -10^{-4}$, the potential is monotonically increasing.
The potential has one local maxima for $\xi= -5\times 10^{-3},
-10^{-3}$ and $-5\times 10^{-4}$. The local maxima shift to the right
as $|\xi|$ decreases. If $|\xi| > 5$, the potential is monotonically decreasing
but we do not plot for this case.}
\label{fig5}
\end{figure}

Similarly, we discuss the following three cases separately

(1) if $\frac{\mu}{\kappa^2 m^2}<|\xi|<\frac{m^2}{\kappa^2 \lambda v^4}$,
there is one local maximum at $\phi=\phi_e$:

(2) if $\frac{\mu}{ \kappa^2 m^2}>|\xi|>0$,
the potential is monotonically increasing:

(3) if $|\xi|> \frac{m^2}{\kappa^2 \lambda v^4}$,
the potential is monotonically decreasing:

\subsubsection*{\rm (1):}

In this case, there is a local maximum in the potential.
Then
inflation can take place

(i) in the vacuum-dominated region where ${\hat V}_{,\phi}>0$

(ii) around the local maximum

(iii) in the large field region

In the region (i),
$\frac{d{\hat V}}{d\phi}\simeq
(m^2-\kappa^2 |\xi|\lambda v^4) \phi >0$.
Thus, $\phi$ rolls down toward the origin.
Then, the slow-roll parameters  and the e-folding number
are given by
Eq. (\ref{eq:slowroll_m0vac})
and Eq. (\ref{eq:efold_m0vac}).
By the same reasoning from Sect. \ref{sect:mu0}
inflation must be the type (I).
The slow-roll inflation can be realized
for
\bea
0<\phi<\phi_i
:=\frac{\kappa\lambda v^4}{2\sqrt{2}(m^2-|\xi|\kappa^2\lambda v^4)},
\eea
where $\epsilon_i=1$.
Noting that $N\simeq \frac{\kappa\phi_i}{\sqrt{2}}$,
in order to realize sufficiently long inflation,
$\phi_i\gg \frac{1}{\kappa}$.
On the other hand,
we also need to require $\phi_e>\phi_i$,
hence
$\phi_e\gg \frac{1}{\kappa}$, which leads to
$$
\frac{m^2}{\kappa^2 \lambda v^4}>|\xi|>\frac{\kappa^2 m^2+\mu}
{\kappa^2 m^2 + \kappa^4 \lambda v^4},
$$
where the upper bound is given from our classification.
Combining with (\ref{eq:efold_m0vac}), we obtain
$$
N< \frac{\kappa^2 m^2 + \kappa^4 \lambda v^4}
     {4(\frac{m^4}{\lambda v^4}-\mu)}.
$$
We have to require $N>50$, hence
$$
\frac{\kappa^2 m^2+ \kappa^4 \lambda v^4}
     {\frac{m^4}{\lambda v^4}-\mu}>200.
$$
For $\mu\lesssim \frac{m^4}{\lambda v^4}$,
this condition can be satisfied.

The amplitude of the curvature perturbations, the tensor-to-scalar ratio,
and the spectral indices of the curvature and tensor perturbations
are given by
\bea
P_{s}&\simeq &\frac{v^{12}\kappa^6 \lambda^3}
  {768 \pi^2\big(m^2-|\xi|\kappa^2\lambda v^4\big)^2\phi_{\ast}^2},
\quad
r\simeq \frac{128(m^2-|\xi|\kappa^2 \lambda v^4)^2\phi_{\ast}^2}
       {v^8\kappa^2\lambda^2},
\nonumber\\
n_{s}-1&\simeq &8\big(\frac{m^2}{\kappa^2 \lambda v^4}-|\xi|\big)>0,\quad
n_t\simeq -\frac{16(m^2-|\xi|\kappa^2 \lambda v^4)^2\phi_{\ast}^2}
          {\kappa^2\lambda^2 v^8}\,.
\eea
Thus, the spectrum of the curvature perturbations is blue tilted.

In the region (ii),
around the local maximum,
the slow-roll parameters are given by
\bea
\epsilon &\simeq&
32(m^2-|\xi|\kappa^2 v^4\lambda)^2(|\xi|\kappa^2 m^2-\mu)^5
(\phi-\phi_{e})^2
\nonumber\\
&\times &
\Big\{\kappa^2
(\kappa^2m^2|\xi|-\mu
+|\xi|\kappa^2(m^2-|\xi|\kappa^2v^4))^2
\nonumber\\
&\times&
\big[
(m^2-|\xi|\kappa^2 v^4\lambda)(1+6|\xi|)\kappa^2|\xi|
+m^2\kappa^2|\xi|-\mu\big]
(m^4-\lambda v^4\mu)^2
\Big\}^{-1}
\nonumber\\
\eta
&\simeq&
-\frac{8(m^2-|\xi|\kappa^2 \lambda v^4)
       (\mu-|\xi|\kappa^2 m^2)^2}
 {\kappa^2
\big[
(m^2-|\xi|\kappa^2 v^4\lambda)(1+6|\xi|)\kappa^2|\xi|
+m^2\kappa^2|\xi|-\mu
\big](m^4-\lambda v^4\mu)}
<0.
\eea
For $\phi<\phi_e$, inflation can be
either the type (I) or the type (II).
For $\phi>\phi_e$, it can be only
the type (II).
The e-folding number is given by
\bea
N
&\simeq &
\frac
{\kappa^2
\big[
(m^2-|\xi|\kappa^2 v^4\lambda)(1+6|\xi|)\kappa^2|\xi|
+m^2\kappa^2|\xi|-\mu
\big](m^4-\lambda v^4\mu)}
{8(m^2-|\xi|\kappa^2 \lambda v^4)
       (\mu-|\xi|\kappa^2 m^2)^2}.
\eea
The spectral index of the curvature perturbations is given by
\bea
n_{s}-1\simeq
-\frac{16(m^2-|\xi|\kappa^2 \lambda v^4)(\mu-|\xi|\kappa^2 m^2)^2}
 {\kappa^2
\big[
(m^2-|\xi|\kappa^2 v^4\lambda)(1+6|\xi|)\kappa^2|\xi|
m^2\kappa^2|\xi|-\mu
\big](m^4-\lambda v^4\mu)}
\simeq -\frac{2}{N}.
\eea
The tensor spectral index and the tensor-to-scalar ratio
are given by
$n_{t}\simeq -2\epsilon$ and
$r\simeq 16\epsilon$.

In the region (iii),
inflation can not be the type (I).
In addition, as $\phi$ increases
the slow-roll parameters
\bea
\epsilon\simeq \frac{8(|\xi|\kappa^2 m^2-\mu)^2}
              {(1+6|\xi|)|\xi|\kappa^4\mu^2\phi^4},
\quad
\eta\simeq \frac{8(|\xi|\kappa^2 m^2-\mu)}
              {(1+6|\xi|)\kappa^2 \mu \phi^2},
\eea
decrease.
Therefore there is no way to terminate inflation.


\subsubsection*{\rm (2):}
The potential is monotonically increasing as $\phi$ becomes larger,
since $\phi_{e}\to \infty$
as $\xi\to -\frac{\mu}{\kappa^2 m^2}$ in the subcase (1).
The inflation can take place

(i) in the vacuum-dominated region

(ii) in the large field region


In the region (i),
$\frac{d{\hat V}}{d\phi}\simeq
(m^2-\kappa^2 |\xi|\lambda v^4) \phi >0$.
Thus, $\phi$ rolls down toward $\phi=0$.
Then, the slow-roll parameters
are given by
\bea
\epsilon
\simeq \frac{8(m^2-|\xi|\kappa^2 \lambda v^4)^2\phi^2}
              {\kappa^2 \lambda^2 v^8},
\quad
\eta\simeq
\frac{8(m^2-|\xi|\kappa^2 \lambda v^4)}
              {\kappa^2 \lambda v^4}>0.
\eea
Since as $\phi$ decreases $\epsilon$ decreases,
inflation cannot be the type (II), but be the type (I).
The e-folding number is given by
\bea
N\simeq \frac{\kappa^2\lambda v^4}
       {4(m^2-\kappa^2 \lambda |\xi|v^4)}.
\eea
The slow-roll inflation can be realized
for
\bea
0<\phi<\phi_i
:=\frac{\kappa\lambda v^4}{2\sqrt{2}(m^2-|\xi|\kappa^2\lambda v^4)},
\eea
where $\epsilon_i=1$.
Noting that $N\simeq \frac{\kappa \phi_i}{\sqrt{2}}$,
to realize sufficiently long inflation,
$\phi_i\gg \frac{1}{\kappa}$, which leads to
$$
\frac{\mu }{\kappa^2 m^2}>|\xi|>
\frac{m^2 }{\kappa^2 \lambda v^4}
-\frac{1}{2\sqrt{2}},
$$
where the upper bound is given from our classifications.

The amplitude of the curvature perturbations, the tensor-to-scalar ratio,
and the spectral indices of the curvature and tensor perturbations are given
by
\bea
P_{s}&\simeq &\frac{v^{12}\kappa^6 \lambda^3}
  {768 \pi^2\big(m^2-|\xi|\kappa^2\lambda v^4\big)^2\phi_{\ast}^2},
\quad
r\simeq \frac{128(m^2-|\xi|\kappa^2 \lambda v^4)^2\phi_{\ast}^2}
       {v^8\kappa^2\lambda^2},
\nonumber\\
n_{s}-1&\simeq &8\big(\frac{m^2}{\kappa^2 \lambda v^4}-|\xi|\big)>0,\quad
n_t\simeq -\frac{16(m^2-|\xi|\kappa^2 \lambda v^4)^2\phi_{\ast}^2}
          {\kappa^2\lambda^2 v^8}\,.
\eea
Thus, the spectrum of the curvature perturbations is blue tilted.

In the region (ii),
$\phi\gg \phi_{e}$,
the slow-roll parameters Eq. (\ref{sl}) are given by
\bea
&&\epsilon
\simeq \frac{8\big(\mu-|\xi|\kappa^2m^2\big)^2}
            {(1+6|\xi|)|\xi|\kappa^4 \mu^2\phi^4},
\quad
\eta
\simeq -\frac{8\big(\mu-|\xi|\kappa^2m^2\big)}
            {(1+6|\xi|)\kappa^2 \mu \phi^2},
\eea
and the slow-roll approximation is valid in the range
\bea
\phi>\phi_f=\frac{2^{3/2}(\mu-|\xi|\kappa^2 m^2)^{1/2}}
{(6|\xi|+1)^{1/2}(\kappa^2\mu)^{1/2}}.
\eea
Inflation can be either the type (I) or the type (I).
If $\phi_c>\phi_f$, inflation is the type (I),
while
if $\phi_c<\phi_f$ it is the type (II).
The e-folding number
from the horizon crossing to the end of inflation
is given by
\bea
N\simeq \frac{(1+6|\xi|)\kappa^2\mu}
       {8(\mu-|\xi|\kappa^2 m^2)}\Big(\phi_\ast^2-\phi_{c,f}^2\Big)
\simeq
\frac{(1+6|\xi|)\kappa^2\mu\phi_{\ast}^2}
       {8(\mu-|\xi|\kappa^2 m^2)},
\eea
where we assume that $\phi_{\ast}\gg\phi_{c,f}$.
The amplitude of the curvature perturbations
and the tensor-to-scalar ratio are given by
\bea
&&P_{s}
\simeq \frac{(1+6|\xi|)\kappa^4 \mu^3\phi_{\ast}^4}
   {768|\xi|\pi^2\big(\mu-|\xi|\kappa^2 m^2\big)^2},
\quad
r\simeq\frac{128(\mu-|\xi|\kappa^2m^2)^2}
       {|\xi| (1+6|\xi|)\kappa^4\mu^2\phi_{\ast}^4}
\simeq \frac{2}{N^2}\big(6+\frac{1}{|\xi|}\big).
\eea
The spectral indices of the curvature and tensor
perturbations become
\bea
&&n_{s}-1
\simeq -\frac{16\big(\mu-|\xi|\kappa^2m^2\big)}
             {(1+6|\xi|)\kappa^2 \mu\phi_{\ast}^2}
\simeq  -\frac{2}{N}<0,
\nonumber\\
&&n_t\simeq
-\frac{16\big(\mu-|\xi|\kappa^2m^2\big)^2}
            {(1+6|\xi|)|\xi|\kappa^4 \mu^2\phi_{\ast}^4}
\simeq
-\frac{1+6|\xi|}{4|\xi|N^2}<0.
\eea
Thus, the spectrum of the curvature perturbations
is red tilted.


We perform the numerical calculations to follow the whole dynamics 
from the local maximum to the vacuum-dominated region.
We consider the cases of $\xi = -5\times 10^{-3}, \,\,
-10^{-3}, \,\,-5\times 10^{-4}$ and $-10^{-4}$ with the parameters
$\lambda = 1,\,\, g^2 = 1, \,\, v^2=10^{-2}m_{pl}^2, \,\, m^2 =10^{-2}m_{pl}^2,
\,\, \mu =10^{-4}$ in Fig. \ref{fig5} in order to compute
the $\phi$ field evolutions, evolution of slow-roll parameters
$\epsilon$ and $\eta$, and the behavior of the spectral index of
the scalar perturbation $n_s$ and the tensor to scalar ratio $r$.
Since the chosen parameter set satisfy the condition $m^4 > \mu
\lambda v^4$, the potential either has a local maxima at $\phi = \phi_e$
or increases as $\phi$ increases. If $|\xi| < \frac{\mu}{\kappa^2 m^2}
\simeq 4\times 10^{-4}$,
the potential increases monotonically as $\phi$ increases ($\phi_e =\infty$).
If $|\xi| \geq \frac{\mu}{\kappa^2 m^2} \simeq
4\times 10^{-4}$, the local maximum locates at $\phi_e$.
We find that as $|\xi|$ increases, the local maximum shifts to
the smaller $\phi$. This implies that if $|\xi| \geq
\frac{m^2}{\kappa^2 \lambda v^4} \simeq 4$,
 the potential
becomes monotonically decreasing function of $\phi$ which will be
discussed below. We choose the initial value of $\phi$ in the
region $\phi_i < \phi_e$ where the $\phi$ field rolls
toward to origin (see Fig. \ref{fig5}-(a)).

Figs. \ref{fig5}-(b) and \ref{fig5}-(c) show the evolution of
the slow-roll parameters $\epsilon$ and $\eta$. 
Since the
slow-roll conditions are not violated, inflation should be
terminated by the tachyonic instability.
Fig. \ref{fig5}-(c) shows negative $\eta$ (negative curvature potential)
for $\xi = -5\times 10^{-3}, \,\, -10^{-3}, \,\, -5\times 10^{-4}$
and positive $\eta$ (positive curvature potential) for $\xi = -10^{-4}$
during slow-roll phase.
We plot $n_s-r$ relation in Fig. \ref{fig5}-(d).
Unlike the case in  Sect. \ref{sect:a}, the spectrum do not cross
to the blue spectrum region ($n_s < 1$) even after the slow-roll phase.

\subsubsection*{\rm (3):}
The potential is monotonically decreasing as $\phi$ becomes larger,
since $\phi_{e}\to 0$,
as $\xi\to -\frac{m^2}{\kappa^2 \lambda v^4}$ in the subcase (1).
In this case,
there is no way to realize the reheating,
since the tachyonic instability does not take place.
Thus, in this paper we do not consider this case.

\subsection{$\frac{m^4}{\mu\lambda v^4}=1$}
\label{sect:c}

\begin{figure}
\centering
\includegraphics[width=0.8\textwidth]{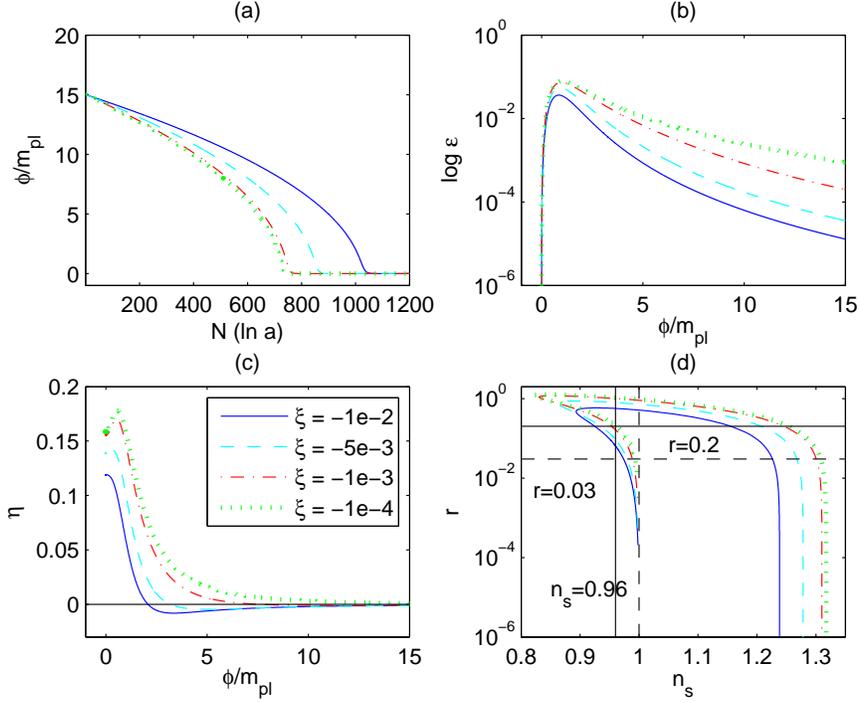}
\caption{We plot same figures as in Fig. \ref{fig1} with
 $\lambda = g^2 = 1, v^2 = 10^{-2}m_{pl}^2, m^2=10^{-4}m_{pl}^2,
\mu = 10^{-4}$ and this parameter satisfy $m^4 = \mu \lambda v^4$
(see Sect. \ref{sect:c}).
For these parameter range, the potential behaves monotonically
increasing or decreasing function of $\phi$.
While the potential decreases as $\phi$ increases for $|\xi| \geq
4\times 10^{-2}$,
the potential is monotonically  increasing
as $\phi$ increases for $|\xi| < 4\times 10^{-2}$ }
\label{fig6}
\end{figure}
\begin{figure}
\centering
\includegraphics[width=0.8\textwidth]{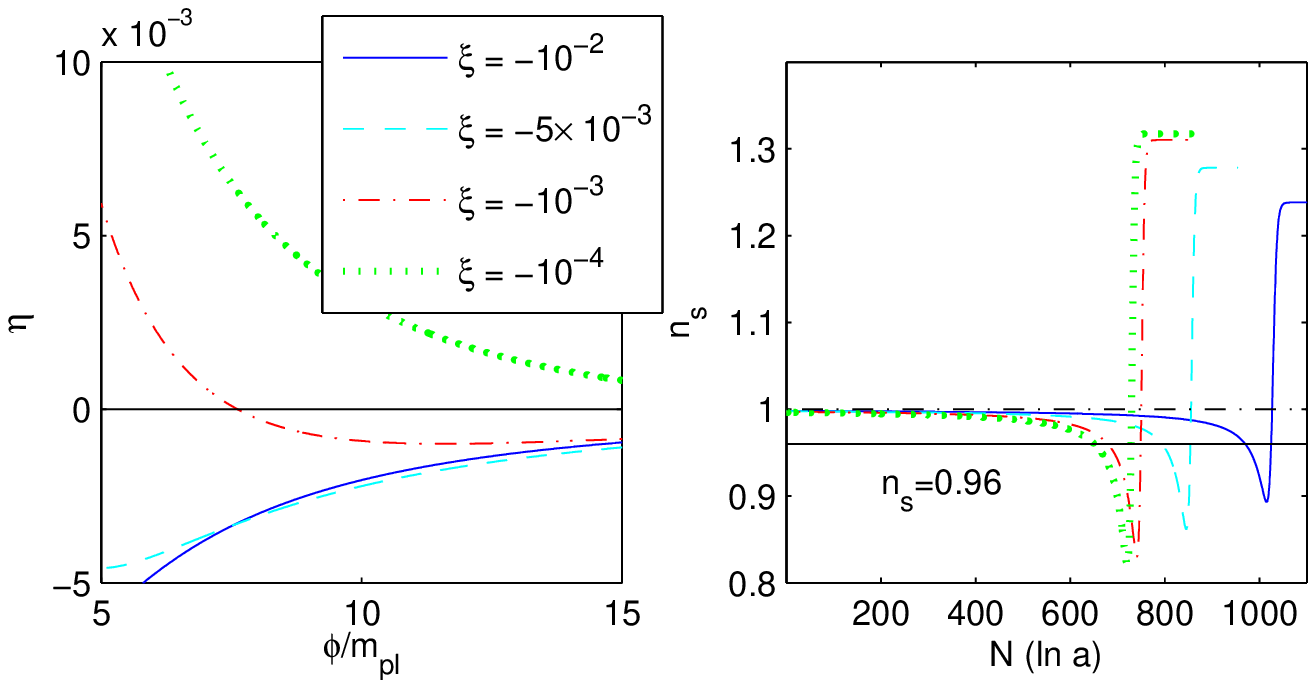}
\caption{ Magnifying the Fig. \ref{fig6}-(c) in the region
$5<\phi < 15$. (right) evolution of spectral index $n_s$
of scalar perturbation.}
\label{fig7}
\end{figure}

In this case, along $\chi=0$
\bea
{\hat V}=\frac{(\lambda v^4+m^2\phi^2)^2}
                    {4\lambda v^4 \big(1+|\xi| \kappa^2\phi^2\big)^2},
\eea
and therefore,
the potential is the monotonic for $\phi$.
Since
$$
\frac{d{\hat V}}{d\phi}
=\frac{(m^2-|\xi|\kappa^4 \lambda v^4)\phi
  (\lambda v^4+m^2\phi^2)}
   {\lambda v^4(1+|\xi| \kappa^2 \phi^2)^3},
$$
for $|\xi|<\frac{m^2}{\kappa^2 v^4\lambda}$,
the potential is increasing,
while
for $|\xi|>\frac{m^2}{\kappa^2 v^4\lambda}$,
it is decreasing.

\subsubsection{$0<|\xi|<\frac{m^2}{\kappa^2 v^4\lambda}$}

Inflation can take place

(i) in the vacuum-dominated region

(ii) in the large field region

In the region (i),
the slow-roll parameters
are given by
\bea
\epsilon
\simeq \frac{8(m^2-|\xi|\kappa^2 \lambda v^4)^2\phi^2}
              {\kappa^2 \lambda^2 v^8},
\quad
\eta\simeq
\frac{4(m^2-|\xi|\kappa^2 \lambda v^4)}
              {\kappa^2 \lambda v^4}>0.
\eea
Since $\epsilon$ is decreasing for decreasing $\phi$,
inflation cannot be the type (II),
but the type (I).
The e-folding number is given by
\bea
N\simeq \frac{\kappa^2\lambda v^4}
       {4(m^2-\kappa^2 \lambda |\xi|v^4)}.
\eea

The slow-roll inflation can be realized
for
\bea
0<\phi<\phi_i
:=\frac{\kappa\lambda v^4}{2\sqrt{2}(-|\xi|\kappa^2\lambda v^4+m^2)},
\eea
where $\epsilon_i=1$.
Noting that $N\simeq \frac{\kappa \phi_i}{\sqrt{2}}$,
to realize sufficiently long inflation,
$\phi_i\gg \frac{1}{\kappa}$, which leads to
$$
\frac{m^2}{\kappa^2 \lambda v^4}
>|\xi|>
\frac{m^2}{\kappa^2 \lambda v^4}
-\frac{1}{2\sqrt{2}}
$$
(if $\frac{m^2}{\kappa^2 \lambda v^4}
>\frac{1}{2\sqrt{2}}$),
where the upper bound is given from our classifications.

The amplitude of the curvature perturbations, the tensor-to-scalar ratio,
and the spectral indices of the curvature and tensor
perturbations are given by
\bea
P_{s}&\simeq &\frac{v^{12}\kappa^6 \lambda^3}
    {768 \pi^2\big(m^2-|\xi|\kappa^2\lambda v^4\big)^2\phi_\ast^2},
\quad
r\simeq \frac{128(m^2-|\xi|\kappa^2 \lambda v^4)^2\phi_\ast^2}
       {v^8\kappa^2\lambda^2},
\nonumber\\
n_{s}-1&\simeq &8\big(\frac{m^2}{\kappa^2 \lambda v^4}-|\xi|\big)>0,\quad
n_t\simeq -\frac{16(m^2-|\xi|\kappa^2 \lambda v^4)^2\phi_{\ast}^2}
          {\kappa^2\lambda^2 v^8}\,.
\eea
Thus, the spectrum of the curvature perturbations is blue tilted.

In the region (ii),
the slow-roll parameters  are given by
\bea
&&\epsilon
\simeq \frac{8\big(m^2-|\xi|\kappa^2\lambda v^4\big)^2}
            {(1+6|\xi|)|\xi|m^4 \kappa^4 \phi^4},
\quad
\eta
\simeq -\frac{8\big(m^2-|\xi|\kappa^2 \lambda v^4\big)}
            {(1+6|\xi|)\kappa^2 m^2 \phi^2},
\eea
and the slow-roll approximation is valid in the range
\bea
\phi>\phi_f=\frac{2^{3/2}(m^2-|\xi|\kappa^2 \lambda v^4)^{1/2}}
{(6|\xi|+1)^{1/2}m \kappa}.
\eea
Inflation can be either the type (I) or the type (II).
If $\phi_c>\phi_f$, inflation is the type (I),
while
if $\phi_c<\phi_f$ it is the type (II).
The e-folding number
from the horizon crossing to the end of inflation
is given by
\bea
N\simeq \frac{(1+6|\xi|)\kappa^2m^2}
       {8(m^2-|\xi|\kappa^2\lambda v^4)}
\Big(\phi_\ast^2-\phi_{c,f}^2\Big)
\simeq
\frac{(1+6|\xi|)\kappa^2\mu\phi_{\ast}^2}
       {8(m^2-|\xi|\kappa^2\lambda v^4)},
\eea
where we assume that $\phi_{\ast}\gg\phi_{c,f}$.
The amplitude of the curvature perturbations
and the tensor-to-scalar ratio are given by
\bea
&&P_{s}
\simeq \frac{(1+6|\xi|)\kappa^4 m^8\phi_{\ast}^4}
   {768|\xi|\pi^2\lambda v^4
\big(m^2-|\xi|\kappa^2 \lambda v^4\big)^2},
\quad
r\simeq\frac{128(m^2-|\xi|\kappa^2\lambda v^4)^2}
       {|\xi| (1+6|\xi|)\kappa^4m^4 \phi_{\ast}^4}
\simeq \frac{2}{N^2}\big(6+\frac{1}{|\xi|}\big).
\eea
The spectral indices of the curvature and tensor
perturbations become
\bea
&&n_{s}-1
\simeq -\frac{16\big(m^2-|\xi|\kappa^2\lambda v^4\big)}
             {(1+6|\xi|)\kappa^2 m^2\phi_{\ast}^2}
\simeq  -\frac{2}{N}<0,
\nonumber\\
&&n_t\simeq
-\frac{16\big(m^2-|\xi|\kappa^2\lambda v^4\big)^2}
            {(1+6|\xi|)|\xi|\kappa^4 m^4\phi_{\ast}^4}
\simeq
-\frac{1+6|\xi|}{4|\xi|N^2}<0.
\eea
Thus, the spectrum of the curvature perturbations
is red tilted.

In order to follow the dynamics of the $\phi$ field 
 from the large field region toward
the vacuum-dominated one, we solve the equations of motion
numerically. 
In Fig. \ref{fig6}, we plot same figures as in Fig. \ref{fig1}
with $\lambda = 1,\,\, g^2 = 1,\,\, v^2 = 10^{-2}\, m_{pl}^2,\,\,
m^2 = 10^{-4}\, m_{pl}^2,\,\, \mu = 10^{-4}$ which satisfy
$m^4 = \mu \lambda v^4$. While the potential decreases monotonically
as $\phi$ increases for  $ |\xi| >  m^2/\kappa^2 \lambda v^4 \simeq
4\times 10^{-2}$, it increases for $|\xi| < 4\times 10^{-2}$.

The evolutions of $\eta$ are shown in Fig. \ref{fig6}-(c) and
the details during slow-roll phase are plotted in Fig. \ref{fig7}-(left).
We plot $n_s-r$ relation for the case of Sect. \ref{sect:c} in
Fig. \ref{fig6}-(d). The evolutions of $n_s$ during slow-roll phase
can be seen in Fig. \ref{fig7}-(right). We find that the spectrum moves
from red to blue.

\subsubsection{$|\xi|>\frac{m^2}{\kappa^2 v^4\lambda}$}

In this case,
there is no way to realize the reheating,
since the tachyonic instability does not occur.
Thus, in this paper we do not consider this case.

\subsection{Summary}

Following the discussions in the previous subsection,
in tables \ref{table_1}, \ref{table_2} and \ref{table_3},
we have summarized the possible inflationary dynamics and
the tilt of the spectrum of the curvature perturbations.
\begin{table}[p]
\caption{\baselineskip 14pt
Classification
in the case $\frac{m^4}{\mu\lambda v^4}<1$.
The type (I) and (II) denote inflation,
terminated by the tachyonic instability and
the violation of slow-roll conditions, respectively.
}
\begin{center}
{\scriptsize
\begin{tabular}{|c||c|c|c|
}
\hline
Coupling parameter&
Region &
Type&
Spectrum
\\
\hline
\hline
$\frac{m^2}{\kappa^2\lambda v^4}
<|\xi|<\frac{\mu}{\kappa^2m^2}$
&
Vacuum-dominated
&
Type (II)
&
Red
\\
\cline{2-4}
&
Local minimum
&
Type (I)
&
Blue
\\
\cline{2-4}
&
Large field
&
Type (I), (II)
&
Red
\\
\hline
$0<|\xi|<\frac{m^2}{\kappa^2\lambda v^4}$
 &
Vacuum-dominated
&
Type (I)
&
Blue
\\
\cline{2-4}&
Large field
&
Type (I), (II)
&
Red
\\
\hline
\end{tabular}
}
\label{table_1}
\end{center}
\end{table}

\begin{table}[p]
\caption{\baselineskip 14pt
Classification
in the case $\frac{m^4}{\mu\lambda v^4}>1$.
Type (I) and (II) denote inflation,
terminated by the tachyonic instability and
the violation of slow-roll conditions, respectively.
}
\begin{center}
{\scriptsize
\begin{tabular}{|c||c|c|c|
}
\hline
Coupling parameter&
Region &
Type &
Spectrum
\\
\hline
\hline
$\frac{\mu}{\kappa^2m^2}
<|\xi|<\frac{m^2}{\kappa^2\lambda v^4}$
&
Vacuum-dominated
&
Type (I)
&
Blue
\\
\cline{2-4}
&
Local maximum
&
Type (I), (II)
&
Red
\\
\hline
$0<|\xi|<\frac{\mu}{\kappa^2m^2}$
 &
Vacuum-dominated
&
Type (I)
&
Blue
\\
\cline{2-4}&
Large field
&
Type (I)
&
Red
\\
\hline
\end{tabular}
}
\label{table_2}
\end{center}
\end{table}

\begin{table}[p]
\caption{\baselineskip 14pt
Classification
in the case $\frac{m^4}{\mu\lambda v^4}=1$.
The type (I) and (II) denote inflation,
terminated by the tachyonic instability and
the violation of slow-roll conditions, respectively.}
\begin{center}
{\scriptsize
\begin{tabular}{|c||c|c|c|
}
\hline
Coupling parameter&
Region&
Type &
Spectrum
\\
\hline
\hline
$0
<|\xi|<\frac{m^2}{\kappa^2\lambda v^4}$
&
Vacuum-dominated
&
Type (I)
&
Blue
\\
\cline{2-4}
&
Large field
&
Type (I), (II)
&
Red
\\
\hline
\end{tabular}
}
\label{table_3}
\end{center}
\end{table}

\section{Conclusions}

In this work, we have investigated the dynamics
and observational consequences in the
hybrid inflationary model
where the inflaton field $\phi$ is nonminimally coupled to
gravity.

In the Jordan frame,
the $\phi^4$ term is added to
the potential of the ordinary hybrid inflation model.
Without this term, the potential in the Einstein frame
decreases in the large field region,
and there is no way to realize the reheating
through the stabilization at the true vacuum.
This new term flattens the potential
in the large field region, and ensures
a sub-Planckian energy density there
for an appropriate choice of parameters.
We have analyzed
the inflationary dynamics
within the typical regions of the potential
in the Einstein frame
in the context of the slow-roll approximations,
and also numerically solved the equations of motion
to investigate the evolution of fields
from the large field region or the local maximum
to the vacuum dominated region or the local minimum.

We have classified inflation into
the type (I) and the type (II).
In these cases, inflation is terminated by
the tachyonic instability and the violation
of the slow-roll approximations,
respectively.
Even in the case of the type (II) inflation,
the tachyonic instability emerges after
the violation of the slow-roll condition,
and the reheating should occur at the true minimum.
Typically, inflation can take place

(1) in the vacuum-dominated region,

(2) around the local maximum,

(3) around the local minimum.

(4) in the large field region,

In the region (1),
inflation
becomes either the type (I) or the type (II),
resulting in the blue or red spectrum of the curvature perturbations,
respectively.
In the region (2),
inflation can be either the type (I) or
the type (II).
They lead to the blue / red spectrum of
the curvature perturbations,
respectively.
In the region (3), inflation must be the type (I),
resulting in the blue spectrum of
the curvature perturbations.
In the region (4), to terminate inflation,
the potential in the Einstein frame must be positively tilted,
which always leads to red spectrum of the curvature perturbations.

We then numerically solved
the equations of motions
from the large field region / the local maximum
to the vacuum dominated region / the local minimum.
The spectrum
of curvature perturbations becomes red tilted
on large scales and
eventually becomes the blue one on smaller scales,
since $\eta$ parameter positively grows
after inflaton field passes through the inflection
point of the potential.
The particularly interesting case is
that inflation starts from the local maximum toward
the vacuum region.
For the optimistic choices of parameters, the spectrum
is always red tilted.

\section*{Acknowledgments}
The authors wish to acknowledge the long-term workshop
``Gravity and Cosmology 2010'' (YITP-T-10-01)
and the YKIS 2010 symposium
``Cosmology --The Next Generation--'',
held at the Yukawa Institute for Theoretical Physics,
during which this work has been initiated.
SK is supported by the National Research Foundation of Korea Grant funded
by the Korean Government [NRF-2009-353-C00007] and
by Basic Science Research Program through the National Research
Foundation of Korea(NRF) funded by the Ministry of Education, Science
and Technology(2010-002596).
MM is grateful for the hospitality of
the Center for Quantum Spacetime, Sogang University.
\appendix



\end{document}